\documentclass[sigconf,natbib=true]{acmart}

\usepackage{booktabs} % For formal tables

\usepackage{url}
\usepackage{epsfig}
\usepackage{bm}
\usepackage{balance}

\copyrightyear{2022}
\acmYear{2022}

\setcopyright{none}
\renewcommand\footnotetextcopyrightpermission[1]{} % removes footnote with conference information in first column
\begin{document}
\fancyhead{}

\title{A Versatile Framework for Evaluating Ranked Lists\\ in terms of Group Fairness and Relevance}

\author{Tetsuya Sakai}
%\authornote{Dr.~Trovato insisted his name be first.}
\affiliation{%
  \institution{Waseda University/Naver Corporation}
%  \streetaddress{}
\city{Tokyo}
\country{Japan}
}
\email{tetsuyasakai@acm.org}

\author{Jin Young Kim}
%\authornote{Dr.~Trovato insisted his name be first.}
\affiliation{%
  \institution{Naver Corporation}
%  \streetaddress{}
\city{Belmont}
\state{CA} 
\country{USA}
}
\email{jin.y.kim@navercorp.com}

\author{Inho Kang}
%\authornote{Dr.~Trovato insisted his name be first.}
\affiliation{%
  \institution{Naver Corporation}
%  \streetaddress{}
\city{Seoul}
\country{Korea}
}
\email{once.ihkang@navercorp.com}

%\renewcommand{\shortauthors}{B. Trovato et. al.}

\begin{abstract}
We present a simple and versatile framework for evaluating ranked lists in terms of group fairness and relevance,
where the groups  (i.e., possible attribute values)  can be either nominal or ordinal in nature.
First, we demonstrate that, if the attribute set is binary,
our framework can easily quantify the overall polarity of each ranked list.
Second, by utilising an existing diversified search test collection and treating each intent as an attribute value,
we demonstrate that our framework can handle soft group membership, and that
our group fairness measures are highly correlated with both adhoc IR and diversified IR measures
under this setting.
Third, we demonstrate how our framework can quantify intersectional group fairness
based on multiple attribute sets. We also show that 
the similarity function for comparing the achieved and target distributions over the attribute values
should be chosen carefully.
\end{abstract}

%\begin{CCSXML}
%<ccs2012>
%<concept>
%<concept_id>10002951.10003317.10003359.10003362</concept_id>
%<concept_desc>Information systems~Retrieval effectiveness</concept_desc>
%<concept_significance>500</concept_significance>
%</concept>
%<concept>
%</ccs2012>
%\end{CCSXML}

%\ccsdesc[500]{Information systems~Retrieval effectiveness}
%\ccsdesc[500]{Information systems~Presentation of retrieval results}

%\keywords{dialogue evaluation; earth mover's distance; evaluation measures; Jensen-Shannon divergence; Kullback-Leibler divergence; order-aware divergence; Wasserstein distance}


\maketitle
%\thispagestyle{empty}

%\newpage

\section{Introduction}\label{s:intro}

Bias can breed bias.
If a ranked list of items presented to the user
is ``unfair'' or ``biased'' from the viewpoint 
of ranked entities (e.g., people, opinions, products, shops) or their stakeholders,
the biased views may influence the users.
Moreover, based on user feedback (e.g., clicks, views) on such ranked lists,
the underlying search engine
may tune itself and further intensify the bias.
Hence, for example, those that are already enjoying much \emph{exposure} or \emph{attention}~\cite{biega18,raj20}\footnote{
While item \emph{exposure} does not necessarily imply user \emph{attention}~\cite{biega18,raj20},
we shall not differentiate between the two hereafter.
}
may further dominate the ranking, 
giving little or no room to those that have never been exposed
to the users.
It is our view that providers of search and ranking services
should strive to prevent and eliminate such vicious circles.
This is our motivation for addressing the problems of  fairness in rankings~\cite{celis17}.

\begin{figure}[t]
\begin{center}

\includegraphics[width=0.45\textwidth]{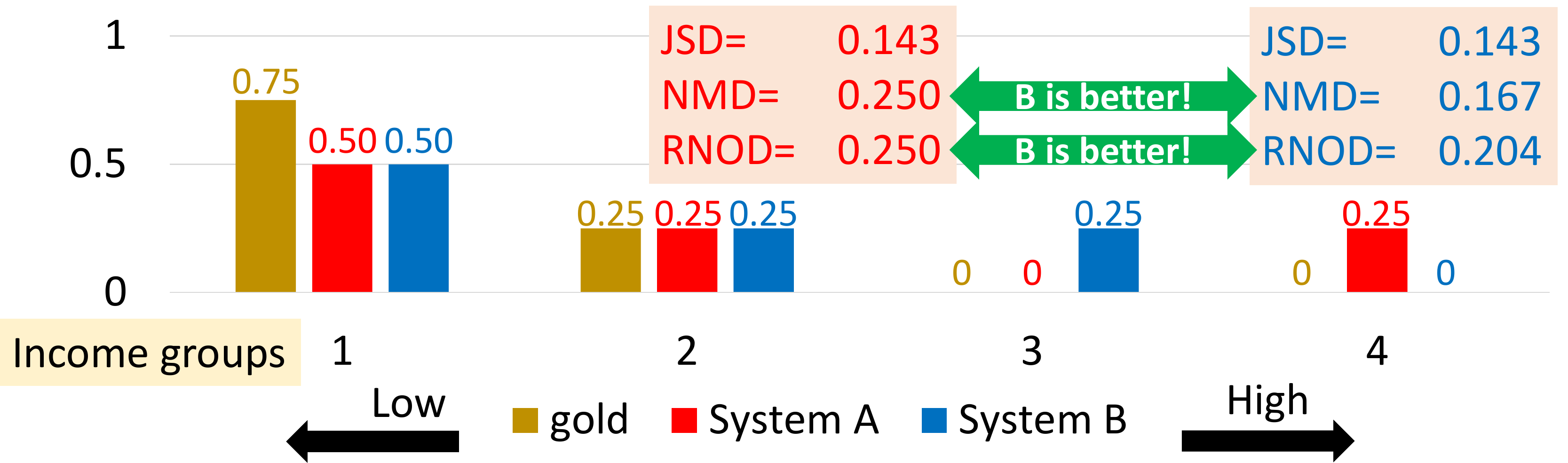}\vspace*{-2mm}
\caption{An example situation where the target distribution emphasises low-income groups
and the distributions achieved by Systems A and B are being evaluated.}\label{f:incomegroups}

\vspace*{-5mm}

\end{center}
\end{figure}

%In the present study,
We address the problem of evaluating ranked lists based on \emph{group fairness}~\cite{ekstrand21},
given a target distribution over \emph{attribute values} in each \emph{attribute set}.
Consider a hypothetical situation where 
we want a group-fair ranking of people, say, scholarship applicants.
Suppose that
the applicants are classified into four classes that represent their income levels (i.e., our attribute values),
and that ideally, we want 75\% of the ranking to represent the lowest income group,
and the other 25\% to represent the second lowest income group.
As the user scans the ranked list of applicants,
the list yields a series of \emph{achieved} distributions over the four classes,
based on the \emph{group membership} of the item (i.e., 
income group of the applicant) at each rank.
To measure the similarity between each achieved distribution and the gold distribution,
%we need a similarity function.
%For this purpose, 
we 
%follow Sakai~\cite{sakai21acl}
%and 
consider utilising \emph{ordinal quantification} measures, namely,
NMD (\emph{Normalised Match Distance}, a normalised version
of \emph{Earth Mover's Distance}~\cite{werman85})
and RNOD (\emph{Root Normalised Order-aware Divergence}~\cite{sakai21acl}),
in addition to a \emph{nominal quantification} measure, namely,
JSD  (\emph{Jensen-Shannon Divergence})~\cite{lin91}.
Ordinal quantification measures take the ordinal nature of the attribute values into account,
and may be appropriate for some applications. For example, 
consider the situation shown in Figure~\ref{f:incomegroups}.
%consider System~A whose achieved distribution over the aforementioned four classes is
%(50\%, 25\%, 0\%, 25\%), and System~B whose achieved distribution is (50\%, 25\%, 25\%, 0\%).
While nominal quantification measures such as JSD say that Systems A and B are equally effective,
NMD and RNOD say that B is better,
as B is leaning more towards the lower income groups (25\% is given to Group~3 rather than Group~4).\footnote{
Calculations based on Eqs.~11, 14, and 19 from Sakai~\cite{sakai21acl}.
}
%Our framework offers these options, unlike previous work in fairness evaluation.

Our main contributions are as follows.
(1)~We present a simple and versatile framework for evaluating ranked lists in terms of group fairness and relevance,
where the groups  (i.e., possible attribute values)  can be either nominal or ordinal in nature.
(2)~We demonstrate that, if the attribute set is binary (e.g., positive vs. negative opinions),
our framework can easily quantify the overall polarity of each ranked list.
(3)~By utilising an existing diversified search test collection and treating each intent as an attribute value,
we demonstrate that our framework can handle soft group membership (i.e., each ranked item 
can have multiple attribute values), and that
our group fairness measures are highly correlated with both adhoc IR and diversified IR measures under this setting.
(4)~We demonstrate how our framework can quantify \emph{intersectional} group fairness~\cite{foulds19} 
based on multiple attribute sets.

\section{Related Work}\label{s:related}

%\subsection{Initial Remarks}\label{ss:initial}

Sections~\ref{ss:NDKL}-\ref{ss:diversity-measures} discuss prior art that are highly relevant to our own group fairness evaluation framework.
Here, we briefly mention prior art that these subsections do not cover.
%
%Before discussing prior art closely related to the present study,
%we briefly mention work on group fairness evaluation measures that are 
%outside our scope.
%First, we will omit the discussion of fairness measures that do not consider a ranking task (e.g.~\cite{speicher18}).
%As for ranked retrieval,
%the group fairness 
The group-fair ranking measures proposed in Kuhlman \textit{et al.}~\cite{kuhlman19}
assume the existence of a gold fair ranking,
and 
%(as opposed to our requirement of a gold
%distribution over attribute values). 
are computed based on
concordant and discordant pairs of ranked items by comparing the gold and system rankings.
%much like Kendall's $\tau$.
Their work focussed on binary attribute sets (i.e., protected vs. non-protected groups).
See also Kuhlman \textit{et al.}~\cite{kuhlman21} for a comparative study of measures under the binary setting.
%
%In the context of measuring the fairness of book recommendation rankings with regards to author gender,
Raj \textit{et al.}~\cite{raj20} compared 
\emph{single-ranking} measures of Yang and Stoyanovich~\cite{yang17} and
of Sapiezynski \textit{et al.}~\cite{sapiezynski19} (which we discuss in Sections~\ref{ss:NDKL} and~\ref{ss:ECE}, respectively)
as well as measures designed for \emph{distributions} and \emph{sequences} of rankings~\cite{biega18,biega20,biega21,diaz20,singh18}.
%, namely,
%the measures of Singh and Joachims~\cite{singh18},
%Biega \textit{et al.}~\cite{biega18},
%Diaz \textit{et al.}~\cite{diaz20},
%and the TREC 2019-2020 Fair Ranking Track measures~\cite{biega20,biega21}.
The latter class of measures is beyond the scope of our work,
as we are interested in evaluating a single ranked list
when the target distribution over attribute values for each attribute set
is given, either top-down 
(e.g., requiring a uniform distribution over all attribute values)
or as a result of some bottom-up derivation
(e.g., requiring \emph{statistical parity}~\cite{ekstrand21}
based on statistics from the target corpus).
%Our framework does not assume, for example,
%that group exposure should be proportional to group relevance,
%although such a requirement can easily be incorporated.
%The measures used at the TREC 2019 Fair Ranking Track~\cite{biega20} were also sequence-based,
%and relied on the assumption that
%group \emph{exposure} (i.e., an overal measure of how much exposure each researcher in a group receives
%in a sequence of academic paper rankings) should be 
%proportional to 
%group \emph{relevance} (i.e., an overall measure of relevance of each researcher's paper for the same group)\footnote{
%This approach has been referred to as \emph{merit-based fairness}~\cite{morik20}.
%}.
%We do not rely on such an assumption.
Also beyond our scope are the following lines of research:
Beutel \textit{et al.}~\cite{beutel19} propose to evaluate group fairness in the context of personalised recommendation
by collecting pairwise item preferences from each user (See also Narasimhan \textit{et al}~\cite{narasimhan20});
Kirnap \textit{et al.}~\cite{kirnap21} estimate
fair ranking measure scores from
\emph{incomplete} group membership labels.

\subsection{Normalised Discounted KL Divergence}\label{ss:NDKL}

Geyik \textit{et al.}~\cite{geyik19} considered two approaches
to evaluating a ranked list of items 
%(e.g., people indexed for LinkedIn Talent Search),
(e.g., people),
where the ranked list is expected to reflect as faithfully as possible 
a given target distribution over
possible \emph{attribute values} (e.g., female, male, other)
in an \emph{attribute set} (e.g., Gender),
or over \emph{combinations} of attribute values from multiple attribute sets
(e.g., Gender AND Age Group).
Let $A=\{a_{i}\}$ denote an attribute set.
Their first proposal is a \emph{set retrieval} measure for the top-$k$ search results,
and is computed for a particular attribute value.
For a given query,
let $p_{\ast}(a_{i})$
denote the \emph{desired} proportion of items having attribute value $a_{i}$ in the ranked list,
s.t. $\sum_{i} p_{\ast}(a_{i}) =1$.
That is, $p_{\ast}$ is the gold probability mass function over $A$.
Let $L$ denote a ranked list,
%to be evaluated, 
and let $L@k$ denote its top-$k$ portion.
Let $p_{L@k}(a_{i})$ be the
\emph{actual} proportion of items having attribute value $a_{i}$
within the top-$k$ results, s.t. $\sum_{i} p_{L@k}(a_{i})=1$.
That is, $p_{L@k}$ is the achieved probability mass function over $A$.
The \emph{Skew} for $a_{i}$ is defined as
$\textit{Sk}(a_{i}, L@k )  = \log_{\mathrm{e}}(p_{L@k}(a_{i})/p_{\ast}(a_{i}))$.
%\begin{equation}\label{eq:skew}
%\textit{Skew}(a_{i}, L@k )  = \log_{\mathrm{e}} 
%(p_{L@k}(a_{i})/p_{\ast}(a_{i})) \ .
%\end{equation}
%Note that ``+1'' does not appear in 
%the original formulation by Geyik \textit{et al.},
%although they do point out that $\log(0)$ should be avoided in practice.
Although 
Geyik \textit{et al.} point out 
that situations where $p_{\ast}(a_{i})=0$ should be avoided,
%However, note that in practice,
note that
$p_{\ast}(a_{i})=0$ may well happen in practice,
especially if combinations of multiple attribute sets are considered 
(e.g., Gender=``male'' AND Age=``$x> 90$'' AND $\ldots$).
They also propose to utilise
%$\max_{a_{i} \in A} {\it Skew}(a_{i}, L@k )$
$\max_{i} {\it Sk}(a_{i}, L@k )$
and 
%$\min_{a_{i} \in A} {\it Skew}(a_{i}, L@k )$
$\min_{i} {\it Sk}(a_{i}, L@k )$
to discuss the quality of the top-$k$ results 
with respect to the attribute set $A$.\footnote{
Similarly,
Pitoura \textit{et al.}~\cite{pitoura17}
proposed  to utilise
%$\max_{a_{i} \in A} |p^{L@k}(a_{i}) - p^{\ast}(a_{i})|$.
$\max_{i} |p_{L@k}(a_{i}) - p_{\ast}(a_{i})|$.
}
However, it is clear that the skew-based measures 
focus only on the worst-case and best-case attribute values.
In summary, skew-based measures  are not adequate
for our purpose because
(a)~they cannot handle ranked retrieval; and
(b)~they do not consider every attribute value in $A$ (when $|A|>2$); and
(c)~$p_{\ast}(a_{i})=0$ can cause inconveniences.

The second proposal by Geyik \textit{et al.} 
%for evaluating group fairness in ranked lists
was to slightly modify a ranked retrieval measure of Yang and Stoyanovich~\cite{yang17}, called \emph{rKL}.
%Although Yang and Stoyanovich~\cite{yang17} considered only binary attribute sets (i.e., protected vs. non-protected),
%they remarked that their measure based on Kullback-Leibler (KL) divergence
%``can be used without modification to go beyond binary protected 
%group membership.''
The modified measure, which Geyik \textit{et al.} refer to as
\emph{Normalised Discounted KL divergence} (NDKL),
utilises
the Kullback-Leibler Divergence (KLD) to compare
the achieved and gold distributions.
%probability mass function with the gold function.
%\begin{equation}\label{eq:ndkl}
%{\it NDKL}(L) = \frac{
%\sum_{k=1}^{|L|} {\it KLD}_{L@k} / \log_{2}(k+1) 
%}
%{
%\sum_{k=1}^{|L|} 1/ \log_{2}(k+1) 
%} \ ,
%\end{equation}
%\begin{equation}\label{eq:KL}
%{\it KLD}_{L@k} = 
%\sum_{i} p_{L@k}(a_{i}) \textit{Skew}(a_{i}, L@k ) \ .
%\end{equation}
\begin{equation}\label{eq:ndkl}
{\it NDKL}(L) = \frac{
\sum_{k=1}^{|L|} \left( \sum_{i} p_{L@k}(a_{i}) \textit{Sk}(a_{i}, L@k ) \right) / \log_{2}(k+1) 
}
{
\sum_{k=1}^{|L|} 1/ \log_{2}(k+1) 
} \ .
\end{equation}
Note that NDKL overcomes Limitations~(a) and~(b)
mentioned above, but not~(c), since it is based on Skew.
Yet another inconvenience with KLD is: (d)~it is unbounded.
We prefer to use 
\emph{Jensen-Shannon Divergence} (JSD) instead as it solves Limitations~(c) and~(d);
Draws \textit{et al.}~\cite{draws21} and the TREC 2021 Fair Ranking Track\footnote{
\url{https://fair-trec.github.io/docs/Fair_Ranking_2021_Participant_Instructions.pdf}
}
has also adopted JSD.
Furthermore, as we have mentioned in Section~\ref{s:intro},
our proposed framework offers 
ordinal quantification measures as possible alternatives to nominal quantification measures such as JSD.
%alternative distribution similarity/divergence functions
%to handle situations such as the one depicted in Figure~\ref{f:4class-example}.
%Note that
%measures based on KLD, JSD, and entropy~\cite{gao20}
%cannot take the ordinal nature of the attribute values into account
%even when this seems appropriate.\footnote{
%Draws \textit{et al.}~\cite{draws21} discuss computing KLD and JSD
%over 7-point scale ratings, but note that the ratings are ordinal.
%}

Geyik \textit{et al.} (and Ghosh \textit{et al.}~\cite{ghosh21,ghosh21arxiv})
argue that
\emph{intersectional} group fairness~\cite{foulds19} can be handled
by considering \emph{combined} attribute values from multiple attribute sets,
such as SkinType AND Gender~\cite{buolamwini18}.
However, we argue that this may not be the best approach to take
if both nominal and ordinal attribute values need to be considered
\emph{and} if it seems appropriate to take the ordinal nature of the attribute values into account.
For example,
consider combining Gender (nominal) and Age Group (ordinal),
and a combined attribute value
Gender=``female'' AND Age=``$x<20$.''
Which of the following two combined attribute values is closer
to the above,
Gender=``male'' AND Age=``$20 \leq x < 40$'' (Gender: \emph{incorrect}; Age Group: \emph{not far off}) or
Gender=``female'' AND Age=``$40 \leq x < 60$'' (Gender: \emph{correct};  Age Group: \emph{far off})?
Similar problems arise when multiple ordinal classes are combined.
Hence, for applications where ordinal quantification measures seem more appropriate 
than nominal ones such as JSD,
we propose to
compare
the achieved and gold distributions 
 \emph{for each attribute set at a time},
and finally aggregate the scores across the multiple attribute sets.

NDKL adopted the log-based discounting scheme of nDCG~\cite{jarvelin02,sakai14promise}
because
``\textit{it is more beneficial for an item to be ranked higher,
it is also more important to achieve statistical parity at higher ranks.}''
The discounting scheme can be interpreted as reflecting user attention over a ranked list.
However,
Ghosh \textit{et al.}~\cite{ghosh21} and Sapiezynski \textit{et al.}~\cite{sapiezynski19}
argue that the log-based decay is too fat-tailed for modelling user attention.
The next section reviews their work.

\subsection{Expected Cumulative Exposure}\label{ss:ECE}

Inspired
by the work of Sapiezynski \textit{et al.}~\cite{sapiezynski19}
that considered user attention over search results,
Ghosh \textit{et al.}~\cite{ghosh21}
presented a group fairness measure called \emph{Attention Bias Ratio} (ABR).
%For a ranked list $L$,
Let $F_{L@k}(a_{i})=1$ if the item at rank~$k$ in ranked list $L$
has attribute value $a_{i}$, and let $F_{L@k}(a_{i})=0$ otherwise.
The \emph{Mean Attention} (MA) score of $a_{i}$ for $L$
is defined as:
\begin{equation}\label{eq:MA}
{\it MA}(a_{i}, L) = \frac{
\sum_{k=1}^{|L|} F_{L@k}(a_{i}) \,\, {\it Attention}_{p}@k
}
{
\sum_{k=1}^{|L|} F_{L@k}(a_{i})
} \ ,
\end{equation}
where
${\it Attention}_{p}@k = 100 p (1-p)^{k-1}$,
which is essentially the decay function of Rank-Biased Precision (RBP)
for a given patience parameter value $\phi=1-p$~\cite{moffat08}.
Note that this decay depends entirely
on the document rank; it considers neither relevance nor fairness of the documents seen so far.
This limitation applies to NDKL (Eq.~\ref{eq:ndkl}) as well.
Hence, if relevance assessments are available, 
we use the cascade-based decay of Expected Reciprocal Rank (ERR)~\cite{chapelle11}
as in Biega \textit{et al.}~\cite{biega20,biega21} and Diaz \textit{et al.}~\cite{diaz20}.

Ghosh \textit{et al.}~\cite{ghosh21} define ABR as:
\begin{equation}\label{eq:attention}
{\it ABR}(L) = 
%\frac{
%\min_{a_{i} \in A} {\it MA}(a_{i}, L) 
%}
%{
%\max_{a_{i} \in A} {\it MA}(a_{i}, L) 
%} \ .
(\min_{a_{i} \in A} {\it MA}(a_{i}, L))/(\max_{a_{i} \in A} {\it MA}(a_{i}, L)) \ .
\end{equation}
Thus, ABR
quantifies the disparity 
between 
the attribute values with lowest and highest mean attention scores.
It is clear that Limitation~(b) mentioned in Section~\ref{ss:NDKL}
applies to this measure as well.
In contrast, our group fairness framework considers \emph{every} attribute value in each attribute set.

In Eq.~\ref{eq:MA}, note that $F_{L@k}(a_{i})$ is a group membership \emph{flag},
representing \emph{hard} group membership.\footnote{
Ghosh \textit{et al.}~\cite{ghosh21arxiv} acknowledge that 
their method ``\textit{does not take into account partial group membership.}''}
However, in the original work of Sapiezynski \textit{et al.}~\cite{sapiezynski19},
group membership is formulated as a probability mass function
over the attribute values (i.e., \emph{soft} group membership).
That is, let $G_{L@k}(a_{i})$ be the \emph{probability} that the item at rank~$k$ in $L$
has attribute $a_{i}$, s.t. $\sum_{i} G_{L@k}(a_{i}) = 1$.
If the group membership probability mass function $G_{L@k}$ is available for each $k$,
then $G_{L@k}(a_{i})$
can replace $F_{L@k}(a_{i})$ in Eq.~\ref{eq:MA}.
For ranked list $L$,
Sapiezynski \textit{et al.}~\cite{sapiezynski19}
compute a probability distribution over $A$
called the \emph{expected cumulative exposure} (ECE),
where 
the probability for $a_{i}$
is given by
\begin{equation}\label{eq:ECE}
E_{L}(a_{i}) = \sum_{k=1}^{|L|} G_{L@k}(a_{i}) \,\, {\it Attention}_{p}@k \ .
\end{equation}
Note that this generalises the numerator of Eq.~\ref{eq:MA}.
Sapiezynski \textit{et al.} propose
to compare $E_{L}$ with the gold probability mass function $p_{\ast}$
%using a distance measure.
%More specifically,
%they assume 
by assuming that both $E_{L}$ and $p_{\ast}$ are binomial distributions,
and conduct a form of statistical significance test with a test statistic threshold
to discuss whether a ranked list is fair or not.
In contrast, we are more interested in quantifying the \emph{degree} of group fairness of a ranked list
rather than binary classification,
and
%(\textit{cf.} \emph{dichotomous thinking} in statistical significance testing~\cite{harlow16,sakai18book}).
do
not rely on any distributional assumptions.
As we shall demonstrate in Section~\ref{ss:diversity},
our framework can also handle soft group membership,
which is important not only for situations
where each ranked item can take multiple attribute values,
but also for situations where the group membership of each item
needs to be \emph{estimated} with some degree of uncertainty.

%Sapiezynski \textit{et al.}~\cite{sapiezynski19} also discuss how
%the gold distribution $p_{\ast}$ can be obtained.
%The first approach is to obtain $p_{\ast}$ from observational data;
%the second is to treat the distribution obtained from the IR system's corpus statistics
%as if it is the desired distribution.
%In our present study, we treat $p_{\ast}$ as a given:
%for example, $p_{\ast}$ could be set based on the actual statistics (e.g., gender ratio),
%or it could even be set to a uniform distribution over $A$.

\subsection{Polarity on a Binary Attribute Set}\label{ss:polarity}

Consider a situation with a single binary attribute set, $A=\{a_{1}, a_{2}\}$, e.g.,
Democrats vs. Republicans~\cite{kulshrestha17,robertson18}.
Suppose that, for each item at rank~$k$ in ranked list $L$,
its \emph{bias score} $b_{L@k}$ is available~\cite{robertson18}, whose range is $[-1,1]$;
a negative score means leaning towards $a_{1}$, a positive score
means leaning towards $a_{2}$, and 0 means neutral.
Kulshrestha \textit{et al.}~\cite{kulshrestha17}
proposed the Output Bias (OB) measure,
which is an average-precision-like measure based on bias scores.
Whereas OB is only applicable to
binary attribute sets,
our framework can handle any number of attribute values.
If bias scores are available~\cite{robertson18} in a binary setting,
%given the bias score for each ranked item,
our framework can leverage them by converting them to 
group membership probabilities as follows.
%and then our proposed measures can be computed as defined in Section~\ref{s:proposed}.
\begin{equation}\label{eq:converting}
G_{L@k}(a_{1}) = (1+b_{L@k})/2 \ , \,\,\, G_{L@k}(a_{2}) = 1- G_{L@k}(a_{1}) \ .
\end{equation}
%Leveraging actual bias scores~\cite{robertson18} with our framework 
Experimental validation of the above method is left for future work.

Gezici \textit{et al.}~\cite{gezici21} also proposed methods 
to quantify bias in SERPs (Search Engine Result Pages) in a binary setting,
where the objective is to achieve \emph{equality of outcome}, i.e., $p_{\ast}(a_{1})=p_{\ast}(a_{2})=1/2$.
They point out that measures like NDKL (Eq.~\ref{eq:ndkl})
cannot tell whether a SERP is biased towards $a_{1}$ and $a_{2}$,
and propose to compute a (weighted) average of the polarity value ($F_{L@k}(a_{1}) - F_{L@k}(a_{2})$) across document ranks,
%\begin{equation}
%F_{L@k}(a_{1}) - F_{L@k}(a_{2}) \ ,
%\end{equation}
where 
$F$ is a group membership flag as before (See Eq.~\ref{eq:MA}),
but returns 1 only when the document at $k$ belongs to the group in question
\emph{and}
 is relevant.
%Thus, only relevant documents contribute to the bias computation.
%Gezici \textit{et al.} consider both 
%an nDCG-like and RBP-like weighting across ranks.
In Section~\ref{ss:touche},
we demonstrate that our framework can easily 
quantify the overall polarity of each ranked list
given a binary attribute set.
%by utilising two extreme target distributions over the binary attribute set.
%More specifically, 
%we show that most of the argument retrieval runs of Touch\'{e} 2020~\cite{bondarenko20}, a CLEF 2020 task,
%are biased towards PRO arguments rather than CON,
%although only relevance-based evaluation was conducted at
%Touch\'{e} 2020.

%While it is true that, for binary attributes,
%measures that compare the gold and achieved distribution
%over attribute values (such as NDKL) cannot 
%quantify the polarity (i.e., whether the ranked list leans towards $a_{1}$ 
%or towards $a_{2}$),
%note that this can easily be investigated
%by comparing the measure scores 
%with two extreme gold distribution settings:
%$a_{1}=1, a_{2}=0$ and $a_{1}=0, a_{2}=1$.
%The score delta according to (say) NDKL
%is indeed a measure of the overall polarity of the ranked list.
%We shall demonstrate this idea using our proposed framework.

\subsection{Diversity Evaluation Measures}\label{ss:diversity-measures}

Cherumanal \textit{et al.}~\cite{cherumanal21}
applied the measures proposed by Yang and Stoyanovich~\cite{yang17}
as well as a diversified search evaluation measure ($\alpha$-nDCG~\cite{clarke11})
to evaluate the Touch\'{e} 2020 argument retrieval runs from CLEF 2020~\cite{bondarenko20};
they observed 
that systems are generally ranked differently by fairness, diversity, and relevance measures.
%they also discuss leveraging the capability of NDKL for handling nonbinary attributes as a future work.
Diaz \textit{et al.}~\cite{diaz20} discussed the connection 
between group fairness measures (for a \emph{distribution} of ranked lists)
and \emph{intent-aware} diversity measures~\cite{agrawal09,chapelle11,sakai14promise}.
On the other hand, for diversity evaluation,
Sakai and Song~\cite[Table~7]{sakai11sigir} demonstrated 
a few advantages of their \emph{D$\sharp$-measures}  over $\alpha$-nDCG
and \emph{intent-aware} measures;
Sakai and Zeng~\cite{sakai20tois}
reported that an instance of the D$\sharp$-measure
called D$\sharp$-nDCG outperformed intent-aware measures
in terms of how the measure agrees with human SERP preferences.

Zehlike \textit{et al.}~\cite{zehlike17} remarked
that D$\sharp$-nDCG can be applied to group-fair ranking evaluation
by treating their binary attribute set (protected/non-protected)
as two search intents behind the same query.
%In our present study, 
We generalise this idea
and use D$\sharp$-nDCG as the representative of existing
diversity measures
in our experiments to see how 
group-fair and diversified ranking evaluations are related
given either hard or soft group membership.
D$\sharp$-nDCG is the average of \emph{intent recall}
(a.k.a. subtopic recall~\cite{zhai03})
and \emph{D-nDCG}~\cite{sakai11sigir};
the difference between the standard nDCG (for adhoc IR)
and D-nDCG (for diversified IR)
is that the latter is based on the \emph{global gain}
of each document $d$:
\begin{equation}\label{eq:gg}
\textit{GG}(d) = \sum_{i \in I_{q}} \textit{Pr}(i \mid q) g_{i}(d) \ ,
\end{equation}
where $I_{q}$ is the set of known intents for topic $q$,
$\textit{Pr}(i \mid q)$ is the probability 
%of Intent~$i$ given the topic,
that a user who enters $q$ as a query has Search Intent~$i$,
and $g_{i}(d)$ is the gain value of $d$ for Intent~$i$.
From Eq~\ref{eq:gg},
it can be observed that
if the intent probabilities are uniform
\emph{and}
the attribute values are mutually exclusive
(i.e., hard group membership is defined),
the global gain reduces to a single gain value 
and therefore that D-nDCG reduces to the standard nDCG.
%Our experiments cover situations with \emph{hard} 
%(i.e., mutually exclusive)
%and \emph{soft} memberships of each ranked item;
%(i.e., each item may belong to one or more attribute values),
%and demonstrate that our framework can handle the latter situation
%just like a diversity measure can. 
This is exactly the situation in our first experiment (Section~\ref{ss:touche}),
as it uses the aforementioned Touch\'{e} 2020 data: each ranked item (i.e., opinion) is
either PRO or CON, but not both.
%We also show that
%our group fairness measures are highly correlated with D$\sharp$-nDCG,
%but that they are not the same.
%This result is in line with that of Cherumanal \textit{et al.}~\cite{cherumanal21}.

\section{Proposed Evaluation Framework}\label{s:proposed}

Our premise is that we are given $M$
\emph{attribute sets},
where the $m$-th attribute set is
 $A^{m}  (m=1, \ldots, M)$
with $i$ specifying a particular \emph{attribute value} $a_{i}^{m} (\in A^{m})$.
For each $A^{m}$,
we are also given a target distribution $p_{\ast}^{m}$ over its attribute values,
s.t. $\sum_{i} p_{\ast}^{m}(a_{i}^{m}) = 1$.
An example setting with $M=2$ would be 
$p_{\ast}^{\mathrm{Gender}}(\mathrm{female})=p_{\ast}^{\mathrm{Gender}}(\mathrm{male})=p_{\ast}^{\mathrm{Gender}}(\mathrm{other})=1/3,
p_{\ast}^{\mathrm{Age}}(x < 20)=p_{\ast}^{\mathrm{Age}}(\mathrm{x \geq 80})=0.2,
p_{\ast}^{\mathrm{Age}}(20 \leq x < 80)=0.6$.
Given these targets, we are also given a ranked list $L$ to evaluate,
where each item at rank~$k$ 
has a group membership probability $G_{L@k}(a_{i}^{m})$
s.t. $\sum_{i} G_{L@k}(a_{i}^{m})=1$ for every $k$.
(Recall that group membership flag $F_{L@k}$ is a special case of $G_{L@k}$.)
If the item at rank~$k$ does not correspond 
to any of the attribute values of $A^{m}$,
we let $G_{L@k}(a_{i}^{m}) = 1/| A^{m} |$.
That is, we assume that the distribution over the attribute values is uniform for that item.
Our objective is to quantify how well $L$ aligns with the target distributions and,
if relevance assessments are also available, evaluate $L$ in terms of both group fairness and relevance.

For each attribute set, we can evaluate the group fairness of $L$ as:
\begin{equation}\label{eq:GF}
\textit{GF}^{m}(L)=
\sum_{k=1}^{|L|} \textit{Decay}_{L@k}^{m} \,\,\, \textit{DistrSim}_{L@k}^{m} \ , 
\end{equation}
where $\textit{Decay}_{L@k}^{m}$ is a function that represents the user attention decay as they go down the ranked list,
and $\textit{DistrSim}_{L@k}^{m}$ compares the achieved distribuiton $p_{L@k}^{m}$ with the target distribution $p_{\ast}^{m}$.
In the present study, we employ the relevance-based decay of ERR~\cite{chapelle09,diaz20} by default,
as we view this user model to be more realistic than those of nDCG and RBP that disregard item relevance.
That is,
\begin{equation}\label{eq:decay}
\textit{Decay}_{L@k} = P_{L@k}^{\mathrm{rel}} \prod_{j=1}^{k-1}(1- P_{L@j}^{\mathrm{rel}}) \,\,\,\,\,\,\,(k>1)
\end{equation}
and $\textit{Decay}_{L@1}=  P_{L@1}^{\mathrm{rel}}$,
where $P_{L@k}^{\mathrm{rel}}=(2^{g}-1)/2^{g}$ if the relevance grade of the item ranked at $k$ is $g$~\cite{chapelle09}.
Note that we have removed the $m$ from Eq.~\ref{eq:decay} as the present study assumes that 
group membership for a particular attribute set does not affect attention decay.
While ``group fairness seen so far'' may well
affect the user attention decay
just like ``relevance seen so far,'' this consideration is left for future work.
When relevance assessments are unavailable, 
we employ the RBP-based decay instead: $\textit{Decay}_{L@k} = (1-\phi)\phi^{k-1}$ with $\phi=0.85$.
This is equivalent to assuming that $P^{\mathrm{rel}}_{L@k}=0.15$ for any $L$ and $k$
when computing Eq.~\ref{eq:decay}.\footnote{
For RBP, we let $\phi=0.85$ because this setting has been shown to align well with users' SERP preferences~\cite{sakai20tois}
\emph{and} was the choice in the recent work of Moffat \textit{et al.}~\cite{moffat17}.
}

For each rank $k$ in $L$, the achieved distribution $p_{L@k}^{m}$ is computed by letting
$p_{L@k}(a_{i}^{m}) = \sum_{j=1}^{k} G_{L@j}(a_{i}^{m})/k$.
This is
the average group membership probability over top~$k$ for attribute value $a_{i}^{m}$.
%Since  $\sum_{i} G_{L@k}(a_{i}^{m})=1$ for each rank~$k$,
%For each attribute value $a_{i}^{m}$,
%let $s_{L@k}(a_{i}^{m})= \sum_{j=1}^{k} G_{L@k}(a_{i}^{m})$ (i.e., group membership probabilities summed across the top $k$ ranks).
%Then, $p_{L@k}(a_{i}^{m}) =
%s_{L@k}(a_{i}^{m}) / \sum_{i} s_{L@k}(a_{i}^{m})$ so that
%$\sum_{i} p_{L@k}(a_{i}^{m}) = 1$.
As for $\textit{DistrSim}_{L@k}^{m}$, 
we consider different functions for comparing $p_{L@k}^{m}$ with $p_{\ast}^{m}$
depending on whether the attribute values are nominal or 
ordinal
and, in the latter case, whether considering the ordinal nature of the attribute values makes sense
(See Section~\ref{s:intro}).
More specifically, given an achieved and the gold probability mass functions $p$ and $p_{\ast}$,
we consider the following options.
%\begin{equation}\label{eq:sim-jsd}
%\textit{DistrSim}^{\mathrm{JSD}}(p \parallel p_{\ast}) = 1 - \textit{JSD}(p, p_{\ast}) \ ,
%\end{equation}
%\begin{equation}\label{eq:sim-nmd}
%\textit{DistrSim}^{\mathrm{NMD}}(p \parallel p_{\ast}) = 1 - \textit{NMD}(p, p_{\ast}) \ ,
%\end{equation}
%\begin{equation}\label{eq:sim-rnod}
%\textit{DistrSim}^{\mathrm{RNOD}}(p \parallel p_{\ast}) = 1 - \textit{RNOD}(p \parallel p_{\ast}) \ ,
%\end{equation}
%where JSD, 
%NMD, 
%and 
%RNOD (which is not symmetric)
%are 
\begin{equation}\label{eq:distrsim}
\textit{DistrSim}(p \parallel p_{\ast}) = 1 - \textit{Divergence}(p \parallel p_{\ast}) \ ,
\end{equation}
where $\textit{Divergence}(p \parallel p_{\ast})$ 
is either JSD, NMD, or RNOD (See Section~\ref{s:intro});
we use notations such as $\textit{DistrSim}^{\mathrm{JSD}}$ where appropriate.
%computed as described in Sakai~\cite{sakai21acl}.
%(JSD and NMD are symmetric but RNOD is not.)
%Note that these measures take the maximum value of 1 when the
%two distributions are maximally different; Eqs.~\ref{eq:sim-jsd}-\ref{eq:sim-rnod} provide \emph{similarity} scores.\footnote{
%A symmetric version of RNOD is available~\cite{sakai21acl}, but symmetry is not needed for our purpose.}
%For the reason discussed in Section~\ref{ss:NDKL},
%either NMD or RNOD should be used for ordinal attributes,
%and JSD should generally be used for nominal attributes.
%However, 
%for \emph{binary} attributes (where $|A^{m}|=2$) which implies that there is no distinction between nominal and ordinal
%attributes, any of the three measures can be used:
%in fact, Sakai~\cite{sakai21cikmlq} points out that
%when $A^{m}= \{ a_{1}, a_{2} \} $, 
%$\textit{NMD}(p, p_{\ast}) = \textit{RNOD}(p \parallel p_{\ast})
%=|p_{1}-p^{\ast}_{1}| =  |p_{2}-p^{\ast}_{2}|$.
%$ holds.
%Hence we have two choices of similarity measures in this case (i.e., JSD or NMD).
Note that
JSD should be used for nominal attribute sets unless the attribute set is binary;
for binary attribute sets,
any of the above divergences can be used
as there is no distinction between nominal and ordinal scales.
In fact,
as Sakai~\cite{sakai21cikmlq} shows that
NMD and RNOD are the same
when $A^{m}= \{ a_{1}, a_{2} \}$,
we have two options for the binary case: JSD and NMD (i.e., RNOD).

Given $M$ attribute sets, we predefine a set of weights $w_{0}, w_{1}, \ldots, w_{M}$ 
s.t. $\sum_{m=0}^{M} w_{m}=1$,
and compute the overall score of $L$ as a weighted average.\footnote{
The aforementioned TREC 2021 Fair Ranking track used a \emph{product} 
of a group fairness score and a relevance-based score, but 
we chose the ability to \emph{weight} the component scores if required.
}
We call it the \emph{GFR (Group Fairness and Relevance)} score.
\begin{equation}\label{eq:GFR}
\textit{GFR}(L) = w_{0} \textit{Relevance}(L) + \sum_{m=1}^{M} w_{m} \textit{GF}^{m}(L) \ ,
\end{equation}
where $\textit{Relevance}(L)$ is a relevance-based score;
we let $w_{0}=0$ if relevance assessments are unavailable.
In either case, the present study only considers unweighted versions of Eq.~\ref{eq:GFR},
and leaves the question of how $w_{m}$'s should be set for future work.
As for the choice of $\textit{Relevance}(L)$,
we consider ERR and \emph{iRBU (intentwise Rank-Biased Utility)}~\cite{sakai20tois,sakai21ecir}
because these measures also rely on the realistic decay function given by Eq.~\ref{eq:decay}
and therefore enable us to rewrite Eq.~\ref{eq:GFR} as follows.
\begin{equation}\label{eq:GFR2}
%\textit{GFR}(L) = 
\sum_{k=1}^{|L|} \textit{Decay}_{L@k} \left( w_{0} \textit{Utility}_{L@k} + \sum_{m=1}^{M} w_{m} \textit{GF}^{m}_{L@k} \right) \ ,
\end{equation}
where $\textit{Utility}_{L@k}=1/k$ for ERR and
$\textit{Utility}_{L@k}=\phi^{k}$ for iRBU.\footnote{
For iRBU,
we let $\phi=0.99$
 in the present study as this setting has been shown to align well with users' SERP preferences~\cite{sakai20tois}.
 This is a parameter inherited from RBP, but is used for computing the SERP utility rather than the %decay~\cite{sakai19sigir}.
decay~\cite{amigo18}.
}
Whereas ERR is suitable for navigational searches,
iRBU is a measure that behaves surprisingly similarly to nDCG~\cite{sakai20tois,sakai21ecir}
and is more geared towards informational searches.
Eq.~\ref{eq:GFR2} implies a user model which says 
that the user is scanning down the ranked list while experiencing a sequence of 
documents with different relevance levels \emph{and}
a gradually changing distribution over the attribute values for each attribute set. 
However, as was mentioned earlier, our current decay component considers relevance only.
%uniform,
%so that GFR reduces to simple averaging of a relevance-based measure score
%and the group fairness scores.

%\begin{figure}

%\caption{.}\label{f:random}

%\end{figure}

%\section{Data}\label{s:data}

\section{Experiments with Real Data}\label{ss:experiments}

This section demonstrates the versatility of our group fairness evaluation framework through three case studies with real data.
Hereafter, we consider evaluating the top 
%$|L|=20$ 
$|L|=10$ 
items of any given ranked list.
All relevance-based measures are computed based on an exponential gain value setting.
That is, a gain value of $2^{g}-1$ is given to each $g$-relevant document ($g=0, 1, 2 \ldots$).
%In addition to GFR scores, we also compute D$\sharp$-nDCG scores 
%using \texttt{NTCIREVAL} toolkit\footnote{
%\url{http://research.nii.ac.jp/ntcir/tools/ntcireval-en.html} (version 200626)
%} 
%(See Section~\ref{ss:diversity}) for comparison wherever 
%the combination of an attribute value and a relevance label
%can be treated as a per-intent relevance label.

\subsection{Ranking Pros and Cons: Quantifying the Polarity with One Binary Attribute Set}\label{ss:touche}

%\subsection{Touch\'{e} 2020 Argument Retrieval Data}\label{ss:touche}

As a case study of a ranking task with a binary attribute set,
we follow Cherumanal \textit{et al.}~\cite{cherumanal21} and utilise 
the Touch\'{e} 2020 
%Argument Retrieval 
Data from CLEF 2020~\cite{bondarenko20}.
The task used the \texttt{args.me} corpus~\cite{ajjour19},
which is a collection of opinions each tagged with either PRO ($a_{1}$)
 or CON ($a_{2}$).
We use Version~1 of \texttt{args.me},\footnote{
See \url{https://webis.de/data/args-me-corpus.html} and\\
\url{https://doi.org/10.5281/zenodo.3274636}}
with the Version~1 qrels file from Touch\'{e} 2020 (containing 2,964 topic-document pairs,
covering 49 topics),\footnote{
The qrels file contains 50 topics, but Topic~25 has no relevant documents.
}
and the 21 submitted runs.
The qrels file offers graded relevance on a 6-point scale:
$g=1,\ldots,5$ along with $-2$ (non-arguments).
We treat the non-argument documents as nonrelevant ($g=0$).
Although the runs were evaluated only in terms of relevance at 
Touch\'{e} 2020,\footnote{
The follow-up task, Touch\'{e} 2021, computed nDCG based on \emph{rhetorical quality}
in addition to that based on relevance~\cite{bondarenko21}.
}
we computed the GFR scores using GF$^{\mathrm{JSD}}$ (group fairness), 
ERR and iRBU (relevance);
hence
the GFR measures are denoted by ERR+GF$^{\mathrm{JSD}}$
and iRBU+GF$^{\mathrm{JSD}}$.\footnote{
We also experimented with GF$^{\mathrm{NMD}}$ (which equals GF$^{\mathrm{RNOD}}$
for binary attributes), but we did not find this benficial over  GF$^{\mathrm{JSD}}$.
}
For now, let us consider a \emph{flat} setting,
where the target distribution is $p_{\ast}(\mathrm{PRO})=p_{\ast}(\mathrm{CON})=0.5$.
In addition,
we also compute D$\sharp$-nDCG
using the \texttt{NTCIREVAL} toolkit\footnote{
\url{http://research.nii.ac.jp/ntcir/tools/ntcireval-en.html} (version 200626)
} 
by treating PRO and CON as two search intents behind a query and
treating the combination of the relevance level and the PRO/CON label for each document 
as a per-intent relevance label.
As for intent probabilities, we also use a uniform distribution: $\textit{Pr}(\mathrm{PRO}\mid q)=\textit{Pr}(\mathrm{CON}\mid q)=0.5$ (See Eq.~\ref{eq:gg}).
%which enables us to directly compare D$\sharp$-nDCG with our measures under the flat setting.
Note that,
as we are dealing with hard group membership in this experiment,
D$\sharp$-nDCG reduces to an average of intent recall and the \emph{standard} nDCG (See Section~\ref{ss:diversity-measures}).

Table~\ref{t:argme-tau} compares the run rankings of
different measures under the flat setting  using Kendall's $\tau$.
%our GFR measures under the flat setting, their component measures,
%and D$\sharp$-nDCG in terms of Kendall's $\tau$.
%It can be observed that:
%\begin{itemize}
%\item[(i)] GF$^{\mathrm{JSD}}$ and GF$^{\mathrm{NMD}}$ are highly correlated ($\tau=0.800$),
%as the only difference between these two measures
%is the $\textit{DistrSim}$ component in Eq.~\ref{eq:GF} (i.e., how the achieved and target distributions are compared).
%\item[(ii)] iRBU and D$\sharp$-nDCG are highly correlated ($\tau=0.867$),
%as it is known that iRBU behaves similarly to nDCG~\cite{sakai19sigir,sakai20tois} (which is equivalent to D-nDCG in this experiment);
%\item iRBU shows higher correlations with the GF measures than ERR does.
%This is because both the $\textit{DistrSim}$ component in Eq.~\ref{eq:GF}
%and iRBU's utility function ($\phi^k$) considers the entire top $k$ results,
%while ERR's utility function ($1/k$) focuses only on the $k$-th document.
%\item[(iii)] Most importantly, GF$^{\mathrm{JSD}}$ and GF$^{\mathrm{NMD}}$
%are only moderately correlated with ERR, iRBU, and D$\sharp$-nDCG,
%which suggests that group fairness-based evaluation
%and relevance-based evaluation are indeed complementary.
%\end{itemize}
It can be observed that:
\begin{itemize}
\item GF$^{\mathrm{JSD}}$ is only moderately correlated with ERR, iRBU, and D$\sharp$-nDCG ($\tau=$0.438-0.667),
which suggests that group fairness evaluation is related to but different 
from adhoc and diversity evaluations, at least
in a hard group membership setting with a binary attribute set.
This high-level observation is in line with 
the results of Cherumanal \textit{et al.}~\cite{cherumanal21} who
also used  the Touch\'{e} 2020 data to
compare NKDL, nDCG, and $\alpha$-nDCG.
\item The GFR measures (i.e., \{ERR, iRBU\}+GF$^{\mathrm{JSD}}$) 
are more highly correlated with D$\sharp$-nDCG ($\tau=$0.800-0.867)
than GF$^{\mathrm{JSD}}$ is.
That is, a combination of a group fairness measure and a relevance measure
is relatively similar to a diversity measure (which actually is the average of intent recall and nDCG in this experiment).
\end{itemize}

\begin{table}[t]
\centering

\caption{System ranking correlations (Kendall's $\tau$ with 95\%CIs) for the 21 Touch\'{e} 2020 runs.
}\label{t:argme-tau}
\begin{scriptsize}

%\begin{tabular}{c|c|c|c|c}
%\hline
%					&GF$^{\mathrm{NMD}}$	&ERR		&iRBU		&D$\sharp$-nDCG\\
%\hline % cutoff20 results
%GF$^{\mathrm{JSD}}$	&0.819					&0.229		&0.524		&0.533\\
%					&[0.686, 0.899]			&[$-$0.081, 0.499]&[0.170, 0.758]	&[0.183, 0.763]\\ 
%GF$^{\mathrm{NMD}}$	& -						&0.124		&0.381		&0.371\\
%					&						&[$-$0.278, 0.489]	&[$-$0.008, 0.670]	&[$-$0.020, 0.664]\\
%ERR					& -						& -			&0.705		&0.657\\
%					&						&			&[0.436, 0.858]	&[0.361, 0.833]\\
%iRBU				& - 						& - 			& - 			&0.838\\
%					&						&			&			&[0.667, 0.925]\\
%					
%\hline
%GF$^{\mathrm{JSD}}$	&0.800					&0.438		&0.667		&0.648\\
%					&[0.655, 0.888]			&[0.154, 0.655]&[0.455, 0.807]&[0.428, 0.795]\\ 
%GF$^{\mathrm{NMD}}$	& -						&0.276		&0.467		&0.486\\
%					&						&[$-$0.031, 0.535]&[0.190, 0.675]&[0.213, 0.688]\\
%ERR					& -						& -			&0.771		&0.790\\
%					&						&			&[0.610, 0.871]&[0.639, 0.882]\\
%iRBU				& - 						& - 			& - 			&0.867\\
%					&						&			&			&[0.764, 0.927]\\					
%\hline
%\end{tabular}\vspace{-5mm}

\begin{tabular}{c|c|c|c|c|c}
\hline
							&ERR			&iRBU			&ERR+GF$^{\mathrm{JSD}}$	&iRBU+GF$^{\mathrm{JSD}}$	&D$\sharp$-nDCG\\
\hline
GF$^{\mathrm{JSD}}$			&0.438			&0.667			&0.552						&0.829						&0.648\\
							&[0.154, 0.655]	&[0.455, 0.807]	&[0.298, 0.733]				&[0.702, 0.905]				&[0.428, 0.795]\\
ERR							& -				&0.771			&0.886						&0.610						&0.790\\
							&				&[0.610, 0.871]	&[0.796, 0.938]				&[0.375, 0.771]				&[0.639, 0.882]\\
iRBU						& -				& -				&0.886						&0.838						&0.867\\
							&				&				&[0.796, 0.938]				&[0.716, 0.910]				&[0.764, 0.927]\\
ERR+GF$^{\mathrm{JSD}}$		& - 				& - 				& -							&0.724						&0.867\\
							&				&				&							&[0.538, 0.843]				&[0.764, 0.927]\\
iRBU+GF$^{\mathrm{JSD}}$		& - 				& - 				& -							& -							&0.800\\
							&				&				&							&							&[0.655, 0.888]\\
\hline
\end{tabular}\vspace{-3mm}

\end{scriptsize}

\end{table}

\begin{figure}[t]
\begin{center}

\includegraphics[width=0.4\textwidth]{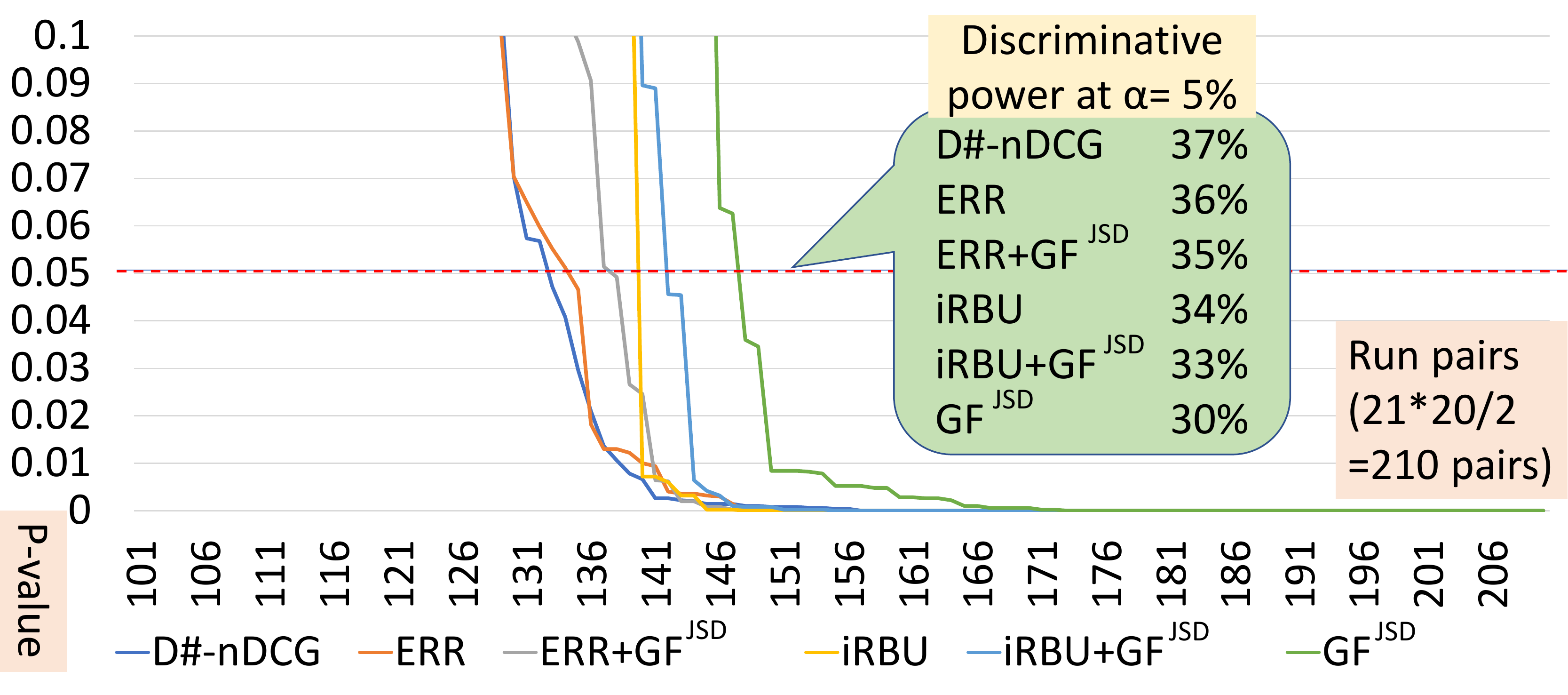}\vspace*{-2mm}
\caption{Discriminative power curves based on the randomised Tukey HSD test (Touch\'{e} 2020 runs).}\label{f:argme-asl}
\vspace*{-5mm}

\end{center}
\end{figure}

%To discuss Observation~(iii) in more detail, 
%we compared the statistical significance test results 
%based on each measure.
Figure~\ref{f:argme-asl} shows the \emph{discriminative power} 
%curves~\cite{sakai06sigir,sakai07sigir,sakai14promise} 
curves~\cite{sakai07sigir,sakai14promise} 
of the measures for the 21 Touch\'{e} runs (210 run pairs)
based on a randomised Tukey HSD test~\cite{sakai18book} with 5,000 trials conducted with the \texttt{Discpower} tool.\footnote{
\url{http://research.nii.ac.jp/ntcir/tools/discpower-en.html}
}
For example, D$\sharp$-nDCG is the most discriminative (with 78 statistically significantly different run pairs out of 210 comparisons, i.e., 37\%) at the 5\% significance level.
When we look at the actual significance test results at the 5\% significance level (i.e., the raw results used to draw Figure~\ref{f:argme-asl}),
a run called \texttt{WeissSchnee-1} is the top performer in terms of D$\sharp$-nDCG: this is the only run that outperforms 7 other runs.\footnote{
This run is also the official top performer of Touch\'{e} 2020 in terms of nDCG~\cite[Table~3(a)]{bondarenko20}.
}
On the other hand,
in terms of statistical significance with ERR, iRBU, and GF$^{\mathrm{JSD}}$,
there are 4, 17, and 14 runs tied at the top;
\texttt{WeissSchnee-1} is in the highest performing cluster for all three measures,
and its rank in terms of mean scores is 
1, 2, and 8, respectively.
That is, 
\texttt{WeissSchnee-1} is only the 8th-best among the 21 runs
in the ranking
according to mean GF$^{\mathrm{JSD}}$.
This example also suggests that
group fairness evaluation is not the same as 
relevance and diversity evaluations,
at least in a hard group membership setting with a binary attribute set.

%However, what is more important is which systems are considered good according to each measure.
%The randomised Tukey HSD test results according to the relevance-based and GF measures 
%tell us the following regarding which runs are the most effective.
%\begin{itemize}
%\item According to D$\sharp$-nDCG,
%WeissSchnee-1 is the winner, as it statistically significantly outperforms the 7 worst performers.
%This is followed by 
%TheThreeMouseketeers-1, PrinceofPersia-\{1,2\}, SwordsmanBaseline-1, and Thongor-1: these runs statistically significantly outperform the 6 %worst performers.
%\item According to ERR,
%the top runs are
%WeissSchnee-1, \\TheThreeMouseketeers-1, PrinceofPersia-1, and
%SwordsmanBaseline-1: they statistically significantly outperform the 6 worst performers.
%\item According to iRBU,
%17 runs constitute the top performing group (which is a superset of the above-mentioned runs), as all of them outperform the  4 worst performers.
%\item  According to GF$^{\mathrm{JSD}}$,
%14 runs constitute the top performing group
%\end{itemize}

\begin{figure}[t]
\begin{center}

\includegraphics[width=0.25\textwidth]{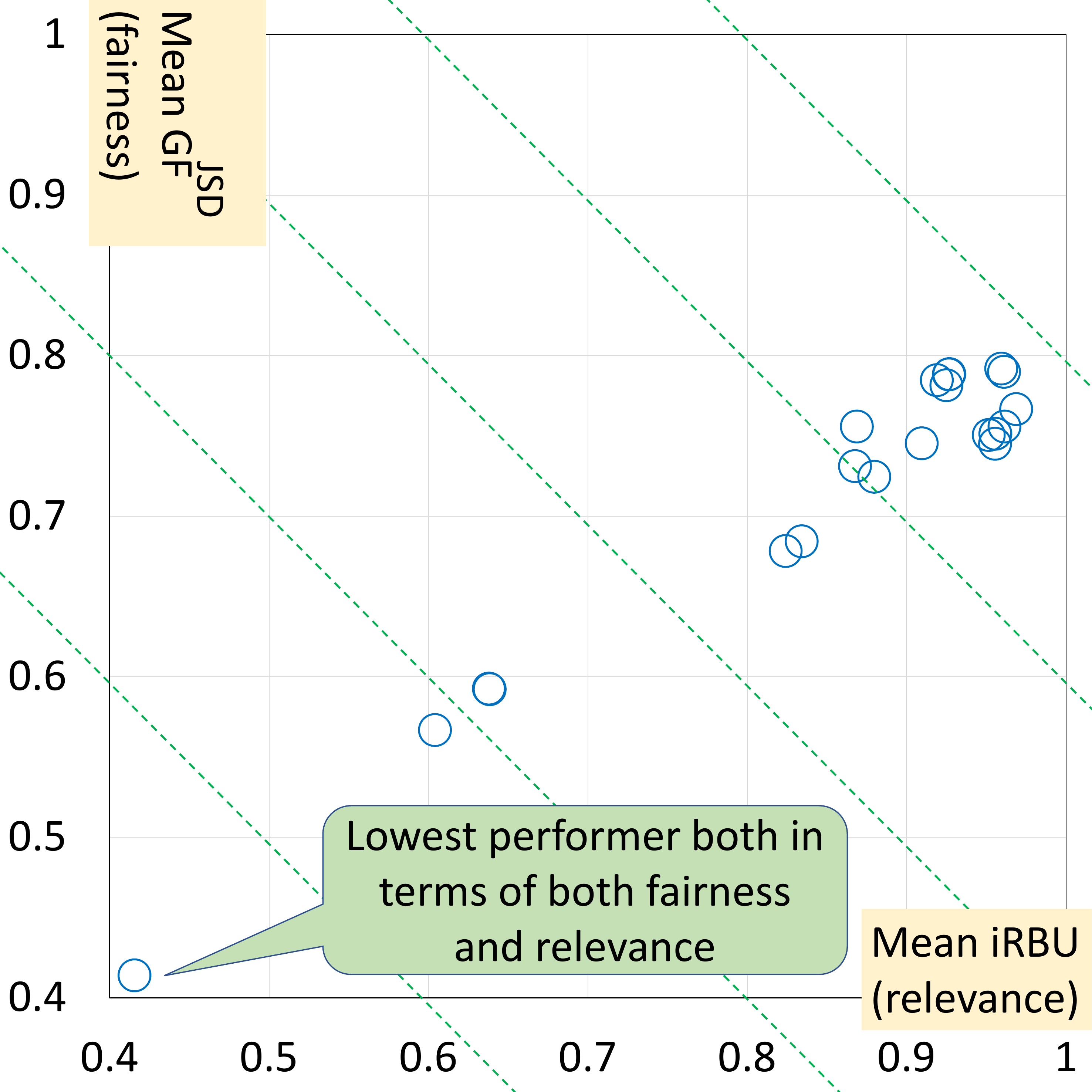}\vspace*{-2mm}
\caption{Visualising the mean group fairness and relevance scores of the 21 Touch\'{e} runs (over 49 topics).}\label{f:argme}

\includegraphics[width=0.25\textwidth]{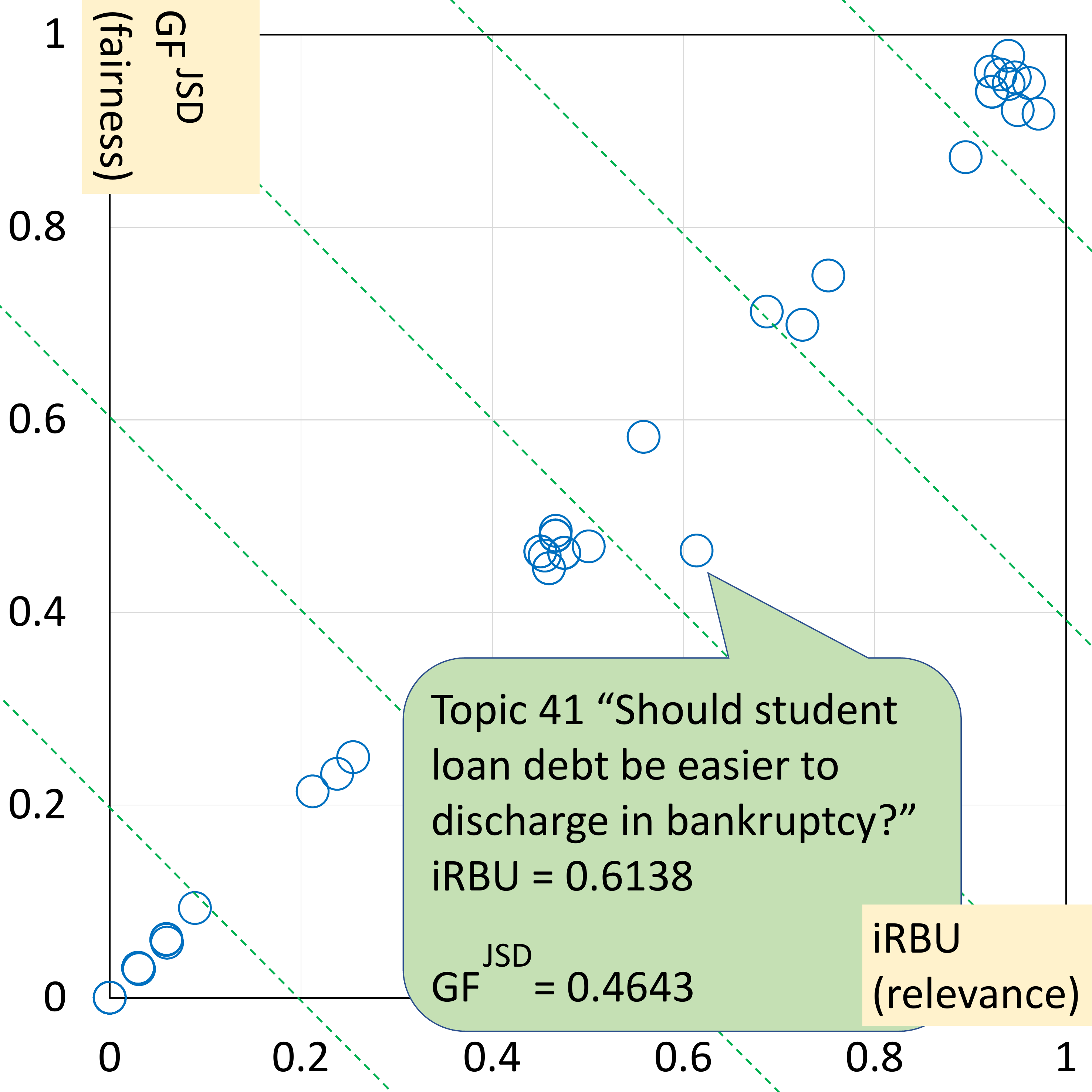}\vspace*{-2mm}
\caption{Visualising the per-topic scores of the lowest performer from Figure~\ref{f:argme}. The Kendall's $\tau$ between iRBU and GF$^{\mathrm{JSD}}$ for this run is 0.867 (95\%CI[0.810, 0.908], $n=49$).
}\label{f:boromir}

\vspace*{-5mm}

\end{center}
\end{figure}

Figure~\ref{f:argme} visualises how the iRBU-based and GF$^{\mathrm{JSD}}$-based run rankings are correlated ($\tau=0.667$ as shown in Table~\ref{t:argme-tau}).
We stress that it is important to
visualise the runs in this way to complement a list of runs ranked by GFR scores (Eq.~\ref{eq:GFR}),
so that we can see how the GF and relevance components are contributing to GFR.\footnote{
Similar practices have been used in the NTCIR INTENT tasks (plotting relevance against diversity)~\cite{song11},
and more recently in the TREC Fair Ranking Tracks (plotting relevance against (un)fairness)~\cite{biega20,biega21}.
}
The dotted lines represent contour lines in terms of GFR (i.e., iRBU+GF$^{\mathrm{JSD}}$).
Figure~\ref{f:boromir} visualises the \emph{per-topic} iRBU and GF$^{\mathrm{JSD}}$ scores
for the lowest performer indicated in Figure~\ref{f:argme}:
as shown with a baloon in Figure~\ref{f:boromir}, we can easily spot SERPs that are relatively poorly balanced between group fairness 
and relevance in this way.

%\begin{table}[t]
%\centering

%\caption{System ranking correlations (Kendall's $\tau$ with 95\%CIs) for the 21 Touch\'{e} 2020 runs: GFR measures
%}\label{t:argme-tau2}
%\begin{scriptsize}
%\begin{tabular}{c|c|c}
%\hline cutoff 20 results
%ERR+GF$^{\mathrm{JSD}}$	&ERR+GF$^{\mathrm{NMD}}$	&0.962 [0.930, 0.980]\\
%iRBU+GF$^{\mathrm{JSD}}$	&iRBU+GF$^{\mathrm{NMD}}$	&0.914 [0.845, 0.953]\\
%\hline
%ERR+GF$^{\mathrm{JSD}}$	&iRBU+GF$^{\mathrm{JSD}}$	&0.619 [0.388, 0.777]\\
%ERR+GF$^{\mathrm{NMD}}$	&iRBU+GF$^{\mathrm{NMD}}$	&0.571 [0.323, 0.746]\\
%
%\hline
%ERR+GF$^{\mathrm{JSD}}$	&ERR+GF$^{\mathrm{NMD}}$	&0.971 [0.946, 0.984]\\
%iRBU+GF$^{\mathrm{JSD}}$	&iRBU+GF$^{\mathrm{NMD}}$	&0.914 [0.845, 0.953]\\
%\hline
%ERR+GF$^{\mathrm{JSD}}$	&iRBU+GF$^{\mathrm{JSD}}$	&0.724 [0.538, 0.843]\\
%ERR+GF$^{\mathrm{NMD}}$	&iRBU+GF$^{\mathrm{NMD}}$	&0.705 [0.510, 0.831]\\
%\hline
%\end{tabular}\vspace{-5mm}
%\end{scriptsize}

%\end{table}

%Table~\ref{t:argme-tau2} compares the system rankings according to different GFR measures in terms of $\tau$.
%The first two lines show that
%the choice of the $\textit{DistrSim}$ function has little impact on the ranking in this experimental setting.
%In contrast, the other two lines show that the choice of the relevance compoment (i.e., ERR or iRBU)
%has a substantial impact on the system ranking.
%Unless the topic set contains many navigational search topics, we recommend iRBU,
%since it is more suitable for informational search intents and is known to behave similarly to the time-honoured nDCG~\cite{sakai20tois}.

We now demonstrate how our framework can quantify the \emph{polarity} of runs, i.e., 
whether the runs are biased towards PRO \emph{or} towards CON, and by how much.
Instead of the flat setting that we considered earlier, 
let us consider a 100\% PRO setting ($p_{\ast}(\mathrm{PRO})=1, p_{\ast}(\mathrm{CON})=0$) and a
100\% CON setting ($p_{\ast}(\mathrm{PRO})=0, p_{\ast}(\mathrm{CON})=1$).
Let $\textit{GF}_{\mathrm{PRO}}(L)$ and $\textit{GF}_{\mathrm{CON}}(L)$ 
denote a GF score for ranked list $L$ computed under the two settings, respectively.
Then $\Delta \textit{GF}(L) = \textit{GF}_{\mathrm{PRO}}(L) - \textit{GF}_{\mathrm{CON}}(L)$
is a direct measure of the polarity of $L$: a positive score implies an overall bias towards PRO, and so on.
Note that replacing the target distribution affects only the $\textit{DistrSim}$ part of Eq.~\ref{eq:GF}.

\begin{figure}[t]
\begin{center}

\includegraphics[width=0.45\textwidth]{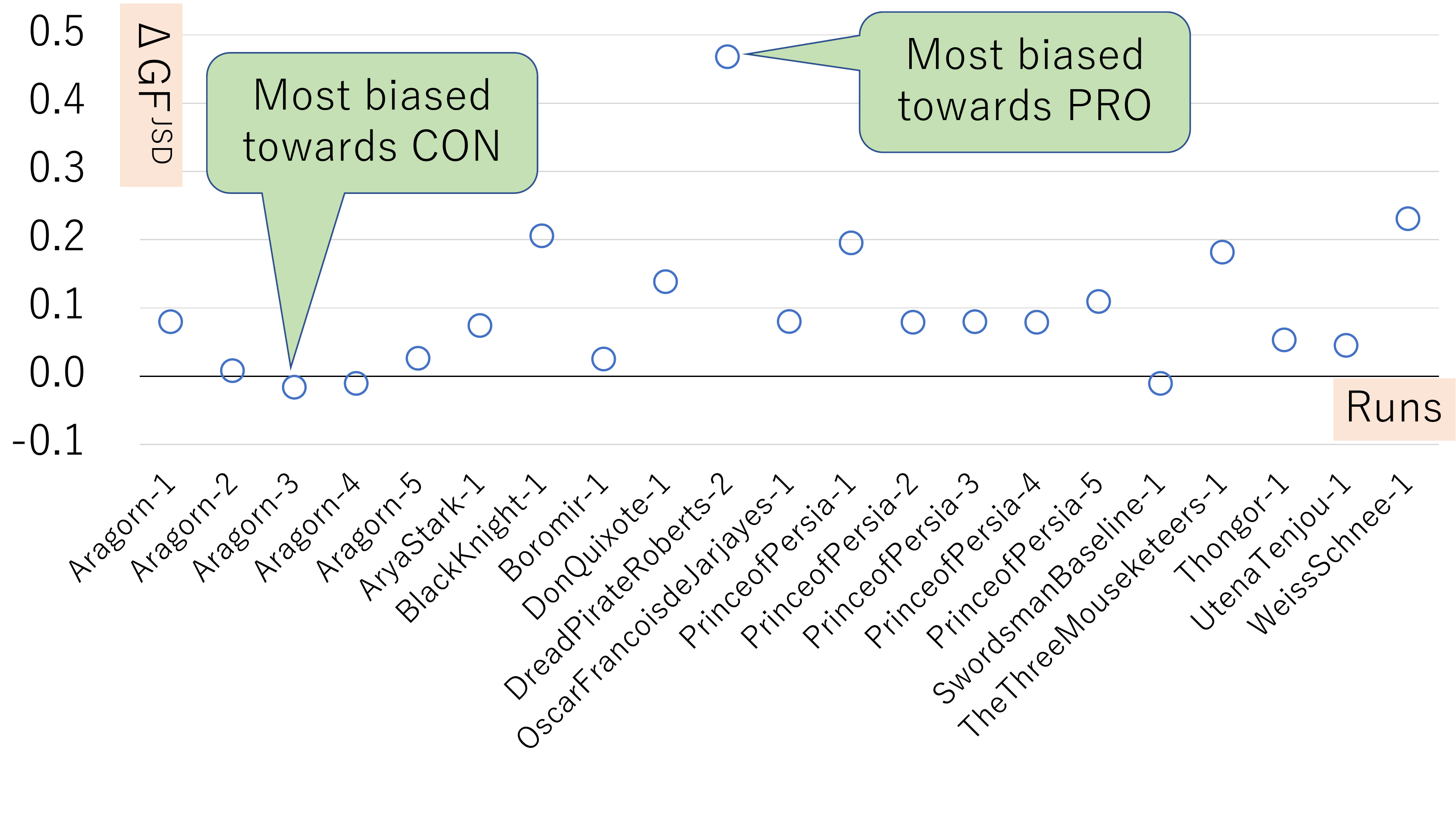}\vspace*{-5mm}
\caption{Quantifying the polarity of the Touch\'{e} 2020 runs ($x$-axis) using Mean $\Delta \textit{GF}$ ($y$-axis) over 49 topics.}\label{f:argme-deltaGF}

\includegraphics[width=0.4\textwidth]{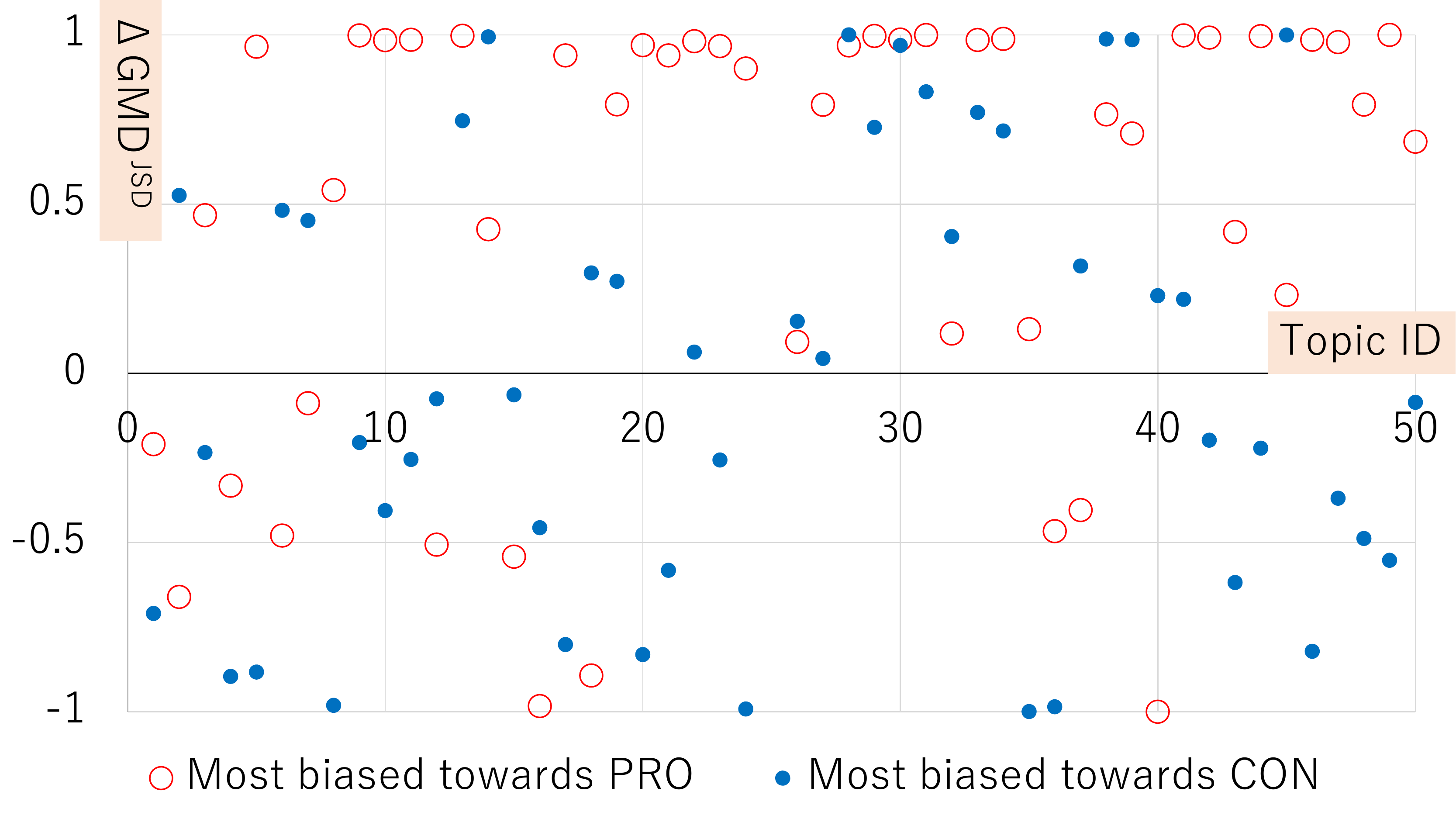}\vspace*{-2mm}
\caption{Per-topic $\Delta \textit{GF}^{\mathrm{NMD}}$ scores for the two most biased runs from Figure~\ref{f:argme-deltaGF}.}\label{f:Team61-Team13}

\vspace*{-5mm}

\end{center}
\end{figure}

Figure~\ref{f:argme-deltaGF} compares the mean $\Delta \textit{GF}^{\mathrm{JSD}}$ scores of the 21 Touch\'{e} 2020 runs.
With the exception of the three runs shown only slightly below the $x$-axis, it can be observed that the runs are
generally biased towards PRO. 
Figure~\ref{f:Team61-Team13} examines the per-topic $\Delta \textit{GF}^{\mathrm{JSD}}$ scores
for the two most extreme runs indicated in Figure~\ref{f:argme-deltaGF}:
it can be observed that the most PRO-biased run is almost completely biased towards PRO for many topics (with 37 topics above the $x$-axis),
while the most CON-biased run is much more well-balanced (with 23 and 26 topics above and below the $x$-axis, respectively).

We have thus demonstrated that our evaluation framework is applicable
to a hard group membership setting with a binary attribute set where
both group fairness and relevance need to be considered,
and that our framework can quantify the polarity of each run as well as each ranked list
in a straightforward manner.

\subsection{Ranking Web Pages: Soft Group Membership 
with One Attribute Set}\label{ss:diversity}

%The Touch\'{e} 2020 experiment described above 
%involved hard group membership, i.e., either PRO or CON.
We now demonstrate that our framework can handle 
soft group membership, i.e., probabilities $G_{L@k}(a_{i})$ rather than flags $F_{L@k}(a_{i})$ like PRO/CON.
To this end, we utilise a diversified search data set 
from the NTCIR-9 INTENT Japanese subtask data~\cite{song11}.
We chose the NTCIR data over the TREC diversity task data~\cite{clarke13}
because  (a)~an NTCIR INTENT task data  set (with associated runs) contains 100 topics
(with 3-24 intents per topic)
while a TREC diversity data set contains only 50;
and more importantly,
(b)~unlike the TREC data, the NTCIR data contains
\emph{intent probabilites} based on assessors' majority votes.
In our experiments, we directly utilise these intent
probabilities to define the target distribution in the group fairness context,
by treating the search intents for each topic as attribute values.
For example, Topic~0127 ``\textit{Mikuniya}'' (a Japanese proper name)
is an ambiguous topic,
because it can be a famous Japanese tea shop,
a restaurant, a hot springs resort, and so on; these are all different entities;
the tea shop intent has a 37\% probability according to the INTENT data,
and we use this directly as the gold probability for group fairness evalution.
Thus,
while search result diversification
aims to satisfy many \emph{users} with different intents behind the query \textit{Mikuniya},
we view the problem in a group fairness context 
where we want to make sure that 
we are giving a fair exposure to each entity named \textit{Mikuniya}.

In the INTENT data,
a single document may be relevant to multiple intents
and therefore soft group membership needs to be handled.
%\footnote{
For example,
for ``\textit{Mikuniya},'' 
there are five documents that are relevant to as many as six intents.
%}
We define the soft group membership for an item based on its per-intent gain values:
if there are three intents and the item has 3, 1, 0 as its per-intent gain values,
the group membership is distributed across the intents as $3/4, 1/4, 0$.
%\footnote
%{
%Even for ambiguous topics like this,
%one document can be relevant to multiple intents and therefore
%soft group membership needs to be handled.
%For example, for Topic~0127, there are five documents 
%that are relevant to as many as six intents.
%}
%for example,
%if the intent probability for Mikuniya the tea shop is 50\%,
%we require the SERP to give this entity a 50\% exposure.
%We thus utilise the INTENT data set 
%to demonstrate how our GFR measures handle soft group membership,
%and how they are related to 
%the official diversity measure from the INTENT task,
%namely, D$\sharp$-nDCG.
%This measure combines the intent probabilities 
%and per-intent graded relevance to compute the global gain 
%of each document~\cite{sakai11sigir}.
%Since we treat each set of intents as a nominal attribute set,
%we use GF$^{\mathrm{JSD}}$ again.
To compute the ERR-based decay (Eq.~\ref{eq:decay}),
we utilise the \emph{per-topic} relevance grades available from Sakai and Zeng~\cite{sakai20tois},
which were derived from the official per-intent relvance assessments.
Along with GF$^{\mathrm{JSD}}$-based GFR measures,
we also compute D$\sharp$-nDCG, the official diversity measure used in the INTENT task.
Recall that D$\sharp$-nDCG utilises the intent probabilities as shown in Eq.~\ref{eq:gg}.

% 18 runs

\begin{table}[t]
\centering

\caption{System ranking correlations (Kendall's $\tau$ with 95\%CIs) for the 18 INTENT runs.
}\label{t:intent-tau}
\begin{scriptsize}

%\begin{tabular}{c|c|c|c}
%\hline
%					&ERR				&iRBU			&D$\sharp$-nDCG\\
%\hline
%GF$^{\mathrm{JSD}}$	&0.843				&0.791			&0.895\\
%					&[0.709, 0.918]		&[0.622, 0.890]	&[0.801, 0.946]\\
%ERR					&-					&0.895			&0.869\\
%					&					&[0.801, 0.946]	&[0.754, 0.932]\\
%iRBU				&-					&- 				&0.817\\
%					&					&				&[0.665, 0.904]\\
%\hline
%\end{tabular}\vspace{-5mm}

\begin{tabular}{c|c|c|c|c|c}
\hline
							&ERR			&iRBU			&ERR+GF$^{\mathrm{JSD}}$	&iRBU+GF$^{\mathrm{JSD}}$	&D$\sharp$-nDCG\\
\hline
GF$^{\mathrm{JSD}}$			&0.843			&0.791			&0.908						&0.882						&0.895\\
							&[0.709, 0.918]	&[0.622, 0.890]	&[0.824, 0.953]				&[0.777, 0.939]				&[0.801, 0.946]\\
ERR							& -				&0.895			&0.935						&0.935						&0.869\\
							&				&[0.801, 0.946]	&[0.874, 0.967]				&[0.874, 0.967]				&[0.754, 0.932]\\
iRBU						& -				& -				&0.882						&0.908						&0.817\\
							&				&				&[0.777, 0.939]				&[0.824, 0.953]				&[0.665, 0.904]\\
ERR+GF$^{\mathrm{JSD}}$		& - 				& - 				& -							&0.974						&0.882\\
							&				&				&							&[0.949, 0.987]				&[0.777, 0.939]\\
iRBU+GF$^{\mathrm{JSD}}$		& - 				& - 				& -							& -							&0.856\\
							&				&				&							&							&[0.731, 0.925]\\
\hline
\end{tabular}\vspace{-3mm}

\end{scriptsize}

\end{table}

\begin{figure}[t]
\begin{center}

\includegraphics[width=0.4\textwidth]{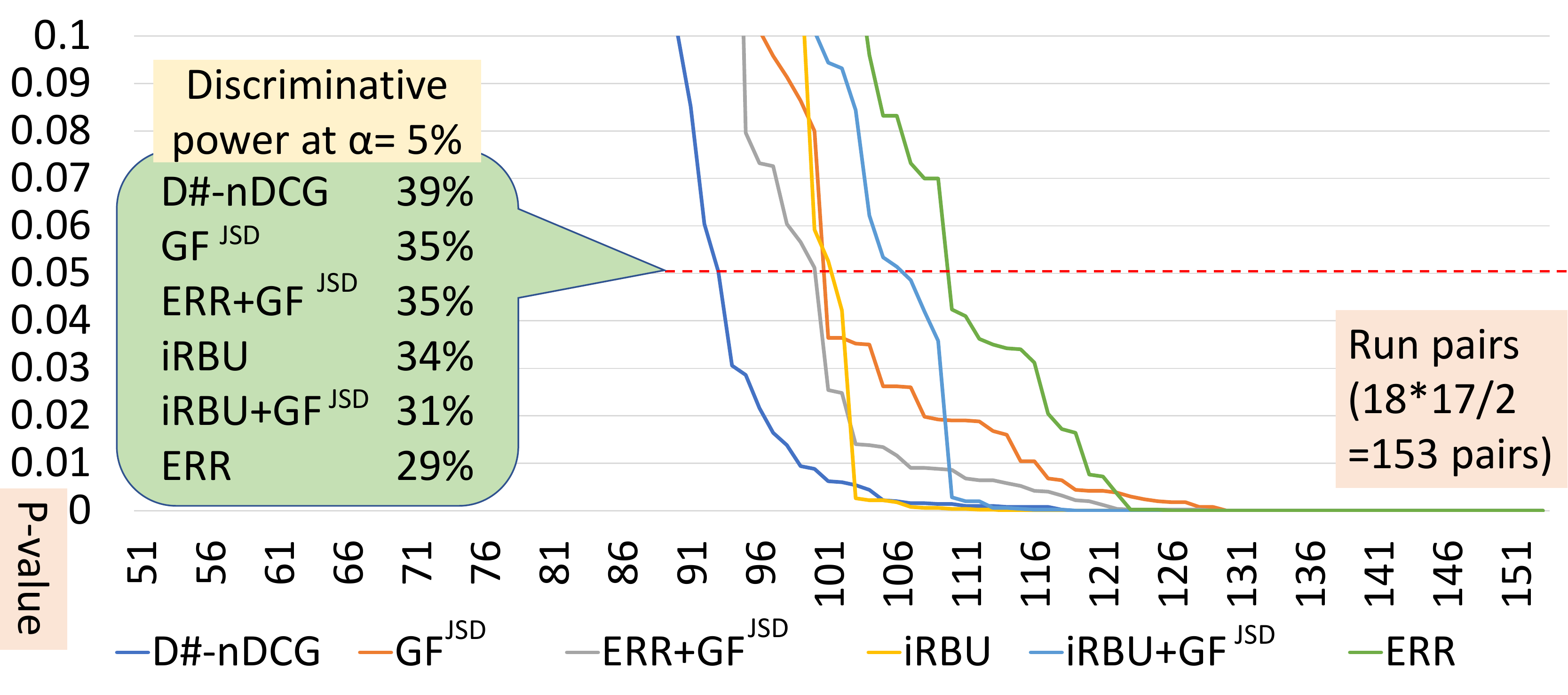}\vspace*{-2mm}
\caption{Discriminative power curves based on the randomised Tukey HSD test  (INTENT runs).}\label{f:intent-asl}
\vspace*{-5mm}

\end{center}
\end{figure}

Using  Kendall's $\tau$,
Table~\ref{t:intent-tau} compares the rankings of the 21 INTENT runs
according to different measures.
Recall that, unlike Table~\ref{t:argme-tau},
we are dealing with soft group membership with 3-24 intents (i.e., attribute values) per topic.
It can be observed that our GF and GFR measures are all highly correlated with 
the offcial diversity measure, i.e., D$\sharp$-nDCG.
Interestingly, GF$^{\mathrm{JSD}}$ is slightly more highly correlated with D$\sharp$-nDCG
than the relevance-based measures (i.e., ERR and iRBU) are, although
the differences are not statistically significant
according to the 95\%CIs.
This suggests that, at least in some search scenarios with soft group membership,
group-fair ranking (for stakeholders of the ranked items) and search result diversification (for search engine users)
may be two sides of the same coin,
or at least, of two similar coins.
This is in contrast to Cherumanal \textit{et al.}~\cite{cherumanal21} and the experiments in Section~\ref{ss:touche}
where hard group membership with a binary attribute set was considered.
One possible cause for the high correlation between GF and D$\sharp$-nDCG
is that we have assumed that the target distribution for group fairness 
(See Eq.~\ref{eq:distrsim})
and the probability distribution of intents given a topic 
(See Eq.~\ref{eq:gg})
are one and the same. In practice, they may well differ,
in which case GF and D$\sharp$-nDCG may possibly give us substantially different results. 

Figure~\ref{f:intent-asl}
shows the discriminative power curves of the measures for the 18 INTENT runs (153 run pairs).
The most discriminative measure is D$\sharp$-nDCG, which is consistent with Figure~\ref{f:argme-asl}.
However, for the INTENT data set, it can be observed that 
GF$^{\mathrm{JSD}}$ is at least as discriminative as ERR and iRBU.
%In terms of statistical significance at $\alpha=0.05$,
When we examine the raw significance test results at $\alpha=0.05$,
D$\sharp$-nDCG says that
four runs (\texttt{MSINT-D-J-\{3,1,2,4\}}) are tied at the top,
statistically significantly outperforming 7 other runs.
As for GF$^{\mathrm{JSD}}$,
it says that \texttt{MSINT-D-J-3} is the top performer,
as it is the only run that statistically significantly outperforms 8 other runs.
In terms of mean scores, all measures except iRBU
agree that  \texttt{MSINT-D-J-3} is the most effective among the 18 runs;
the ranking by mean iRBU
says that this run is the third best.

We have thus demonstrated that our framework can handle soft group membership,
and that GF$^{\mathrm{JSD}}$ (group fairness) and D$\sharp$-nDCG (diversity and relevance)
are highly correlated under this setting.

%\subsection{Ranking Local Shops and Restaurants: Intersectional Group Fairness with
%Nominal and Ordinal Attribute Sets}\label{ss:naver}

\subsection{Ranking Local Shops and Restaurants: Intersectional Group Fairness}\label{ss:naver}

Our third case study involves \emph{two} attribute sets,
one with nominal attribute values and the other with ordinal attribute values (but without relevance data).
The purpose of this experiment is to show that
(i)~unlike prior art, our framework can consider the ordinal nature of the attribute values if required; 
(ii)~for handling ordinal attribute values, the choice of the 
\textit{DistrSim} function (Eq.~\ref{eq:distrsim}) matters in some cases; and
(ii)~intersectional group fairness of rankings can be examined without directly 
combining different attribute sets.

For this experiment, we constructed a data set based on a query log from a popular Local Shop and Restaurant Search service
for smartphone users in Japan.
Given a query, this Local Search service returns a ranked list of items (i.e., shops and restaurants)
based on various features including the relevance to the query, the proximity of the item to the user's location,
and user ratings (i.e., review scores).
In our query log, each ranked item 
has a flag indicating whether it is a \emph{chain store} owned by a \emph{company},
and if it is, the name of the company that owns it.
For some queries, a small number of companies that own many chain stores may
dominate the ranking.
Hence, as an example of imposing a group fairness requirement based on a nominal attribute set (with hard group membership),
we require, \emph{for each query}, that the ideal ranking should provide the same exposure to 
all relevant companies. More specifically,
the gold distribution is defined as a uniform distribution over all companies that appear in the top 20 ranking (based on the current Local Search results)
for that query, where
each shop or restaurant that is \emph{not} a chain store is treated as a distinct company.
In order to demonstrate that our framework can evaluate group fairness from the above viewpoint,
we first obtained a random sample from a one-year Local Search query log (from September 2020 to September 2021),
and then filtered it so that each ranking 
contains at least one company with multiple chain stores listed in the top 20.
This gave us a set of 418 queries for our experiment.

Our query log also contains a mean 5-point scale user rating score and a \emph{review count} for each item.
If an item has $n(>0)$ reviews, the \emph{mean rating} is the average over $n$ user ratings;
if there is no review, the mean rating is set to zero.
We believe that imposing a group fairness requirement based on
review count is of practical significance,
 because this statistic probably reflects the level of exposure of each item in the past, 
and items with low past exposure may deserve more future exposure.
Hence, as an example of handling an ordinal attribute set (with hard group membership),
 we consider a group fairness constraint based on this view.
In the aforementioned query log (before filtering by chain store information),
45\% of the ranked items had zero reviews (Group~1),
22\% had 1-10 reviews (Group~2),
23\% had 11-100 reviews (Group~3),
and the remaining 10\% had over 100 reviews (Group~4).
We utilise this distribution as the gold distribution over the four review count groups \emph{for all queries},
to demonstrate how this search application can be evaluated in terms of \emph{statistical parity}~\cite{ekstrand21}.\footnote{
The exact probabilities we used in our calculations (based on the statistics from our query log) are:
$p_{\ast}(a_{1})=0.452239, p_{\ast}(a_{2})=0.220319, 
p_{\ast}(a_{3})=0.227721, p_{\ast}(a_{4})=0.0997214$.
}

We evaluate the following four runs (i.e., ranking schemes) to demonstrate
how our framework can handle a nominal attribute set and an ordinal attribute set at the same time
and thereby
enable us to quantify intersectional group fairness.
\begin{description}
\item[Base] This represents the actual rankings returned by our current Local Search engine.
%This takes into accout factors such as relevance to the query, user proximity, and user ratings,
The search engines leverages various features to produce the rankings as mentioned earlier, 
but it suffices to treat it as a black box for the purpose of the present experiment. 
\item[Rating] This reranks the top 20 search results by the mean ratings from user reviews,
so that items with higher ratings are prioritised. 
This is likely to affect the review count-based group fairness.
For example, \emph{if} many of the items with higher ratings are those 
that have already enjoyed good exposure to users and therefore received many reviews
(i.e., there is a high correlation between mean rating and review count),
\textbf{Rating} may underperform \textbf{Base} in terms of review count-based group fairness:
recall that our gold distribution for review count groups 
allocates 45\% to Group~1 (items with zero reviews).
The actual Kendall's $\tau$ between the mean ratings 
and the review counts based on all ranked items for the 418 queries ($n=9,329$ items)
is 0.617 (95\%CI[0.609, 0.625]); hence 
there is indeed a substantial positive correlation.
On the other hand, note that
even an item with only one review can have a mean rating of 5 (i.e., maximum);
that is, ranking by mean rating is \emph{not} the same as ranking by review count.
\item[Base-UC] Filter the top 20 items of the \textbf{Base} run
so that each company (which can own multiple chain stores) appears no more than once. UC stands
for ``Unique Chain.'' This is designed to \emph{improve} the chain-based group fairness. 
%Recall that we set up a uniform gold distribution over all companies involved for each topic.
\item[Rating-UC] Similar to \textbf{Base-UC}, except that the input to the filtering step is the \textbf{Rating} run.
This should also improve the chain-based group fairness of \textbf{Rating}.
\end{description}

Since relevance assessments are missing in this experiment,
we use the RBP decay (See Section~\ref{s:proposed}).
For discussing the chain-based group fairness (where the number of nominal bins for the uniform gold distribution varies across topics),
we compute GF$^{\mathrm{JSD}}$.
%note that NMD and RNOD should not be used for nonbinary nominal attributes.
For discussing the review count-based
group fairness (with a statistical parity-based gold distribution over 4 ordinal bins common to all topics),
we compute GF$^{\mathrm{JSD}}$, GF$^{\mathrm{NMD}}$, and GF$^{\mathrm{RNOD}}$.
We shall denote these measures
as chain-GF$^{\mathrm{JSD}}$, revcnt-GF$^{\mathrm{NMD}}$, etc.;
we also average a chain-based score and a revcnt-based score
as a special case of Eq.~\ref{eq:GFR} with $w_{0}=0$ (i.e., doing without relevance)
and $w_{1}=w_{2}=0.5$ (where $i=1,2$ represent the chain and revcnt attribute sets, respectively).
We shall refer to the averages as \emph{intersectional measures} and denote them by
chain-GF$^{\mathrm{JSD}}$+revcnt-GF$^{\mathrm{NMD}}$ and so on.

\begin{table}[t]
\centering

\caption{Mean GF scores of the 4 Local Search runs (over 418 topics).
}\label{t:naver-mean}
\begin{scriptsize}
\begin{tabular}{c|c|c|c|c}
\hline
			&chain-GF$^{\mathrm{JSD}}$	&revcnt-GF$^{\mathrm{JSD}}$	&revcnt-GF$^{\mathrm{NMD}}$	&revcnt-GF$^{\mathrm{RNOD}}$\\
\hline			
Base		&0.404 						&0.495 						&0.542 						&0.460\\
Base-UC		&0.479 						&0.545 						&0.575 						&0.500\\
Rating		&0.401 						&0.445 						&0.551 						&0.422\\
Rating-UC	&0.479 						&0.506 						&0.584 						&0.470\\
\hline
\end{tabular}
\end{scriptsize}

\caption{Randomised Tukey HSD test results ($\alpha=0.05$) for Table~\ref{t:naver-mean}.
``$\gg$'' means ``statistically significantly better than.''
%All the $p$-values for the differences mentioned below are actually smaller than 0.003.
}\label{t:naver-tukey}
\begin{scriptsize}
\begin{tabular}{l|c}
\hline
Measure	&Conclusions\\
\hline
\multicolumn{2}{l}{(a) 4 pairs with a statistically significant difference}\\
\hline
chain-GF$^{\mathrm{JSD}}$,			&\textbf{Rating-UC} $\gg$ \textbf{Base}, \textbf{Rating}\\
revcnt-GF$^{\mathrm{NMD}}$,			&\textbf{Base-UC} $\gg$ \textbf{Base}, \textbf{Rating}\\
chain-GF$^{\mathrm{JSD}}$+revcnt-GF$^{\mathrm{NMD}}$	&\\
\hline
\multicolumn{2}{l}{(b) 5 pairs with a statistically significant difference}\\
\hline
revcnt-GF$^{\mathrm{JSD}}$,			&\textbf{Base-UC} $\gg$ \textbf{Rating-UC}, \textbf{Base}, \textbf{Rating}\\
revcnt-GF$^{\mathrm{RNOD}}$			&\textbf{Rating-UC} $\gg$ \textbf{Rating}\\
									&\textbf{Base} $\gg$ \textbf{Rating}\\
\hline
\multicolumn{2}{l}{(c) 6 pairs with a statistically significant difference}\\
\hline
chain-GF$^{\mathrm{JSD}}$+revcnt-GF$^{\mathrm{JSD}}$,	&\textbf{Base-UC} $\gg$ \textbf{Rating-UC}, \textbf{Base}, \textbf{Rating}\\
chain-GF$^{\mathrm{JSD}}$+revcnt-GF$^{\mathrm{RNOD}}$	&\textbf{Rating-UC} $\gg$ \textbf{Base}, \textbf{Rating}\\
													&\textbf{Base} $\gg$ \textbf{Rating}\\	
\hline
\end{tabular}\vspace{-5mm}								
\end{scriptsize}

\end{table}

Table~\ref{t:naver-mean} shows the mean GF scores of the 4 runs averaged over the 418 topics;
intersectional measures are omitted here as they can easily be computed from the table.
Table~\ref{t:naver-tukey} summarises the significance test results for all 7 measures.
Due to the large sample size ($n=418$), the differences shown in the table
are all statistically highly significant ($p<0.003$). We can observe that:
\begin{itemize}
\item[(I)] With all 7 measures,
\textbf{Base-UC} statistically significantly outperforms \textbf{Base},
and 
\textbf{Rating-UC} statistically significantly outperforms \textbf{Rating}.
That is, the Unique Chain filtering step improves not only the chain-based GF 
but also the revcnt-based GF. Put another way,
trying to give each company a chance (regardless of how many chain stores they own)
also diversifies the review counts in the rankings.
\item[(II)] In terms of chain-GF$^{\mathrm{JSD}}$,
\textbf{Base} and \textbf{Rating} are equally effective.
In other words, 
reranking by mean rating has a negligible effect 
on how different companies are represented in the rankings.
\item[(III)] Interestingly,
while revcnt-GF$^{\mathrm{JSD}}$ and revcnt-GF$^{\mathrm{RNOD}}$
say 
that \textbf{Base} statistically significantly outperforms \textbf{Rating} (Table~\ref{t:naver-tukey}(b)),
revcnt-GF$^{\mathrm{NMD}}$ says that \textbf{Base} slightly \emph{underperforms} \textbf{Rating}
on average. (The difference is not statistically significant).
The results with revcnt-GF$^{\mathrm{JSD}}$ and revcnt-GF$^{\mathrm{RNOD}}$
seem more intuitive,
since 
mean ratings are highly correlated 
with review counts and therefore
\textbf{Rating} probably tends to promote items with high review counts:
at least, it is sure to demote items with zero reviews,
since zero-review items are treated as zero-rating items in our experiments.
\item[(IV)] Two of our intersectional measures found extra
statistically significant differences compared to the component GF measures (Table~\ref{t:naver-tukey}(c)).
This demonstrates that our approach to handling intersectional group fairness effectively
leverages information from both group fairness requirements.
For example, while the difference between \textbf{Rating-UC} and \textbf{Base} 
is not statistically significant in terms of revcnt-GF$^{\mathrm{RNOD}}$  (Table~\ref{t:naver-tukey}(b)),
it is statistically significant in terms of 
chain-GF$^{\mathrm{JSD}}$+revcnt-GF$^{\mathrm{RNOD}}$  (Table~\ref{t:naver-tukey}(c)).
%This reflects the improvements in two directions:
%the reranking by rating which improves the Review Count-based group fairness,
% and the Unique Chain filtering which improves the Chain-based group fairness.
\end{itemize}
Observation~(III) is of particular importance, as this means that the choice of the \textit{DistrSim} function (Eq.~\ref{eq:distrsim}) matters in some cases.
%To the best of our knowledge, this has never been considered in prior art in the field of group fairness evaluation.
Recall that while NMD and RNOD take the ordinal nature of the attribute values into account, JSD cannot;
yet, regarding the comparison of \textbf{Base} and \textbf{Rating},
it is actually revcnt-GF$^{\mathrm{NMD}}$ that is the outlier among the three revcnt-GF measures.
%revcnt-GF$^{\mathrm{JSD}}$ and revcnt-GF$^{\mathrm{RNOD}}$

\begin{figure}[t]
\begin{center}

\includegraphics[width=0.48\textwidth]{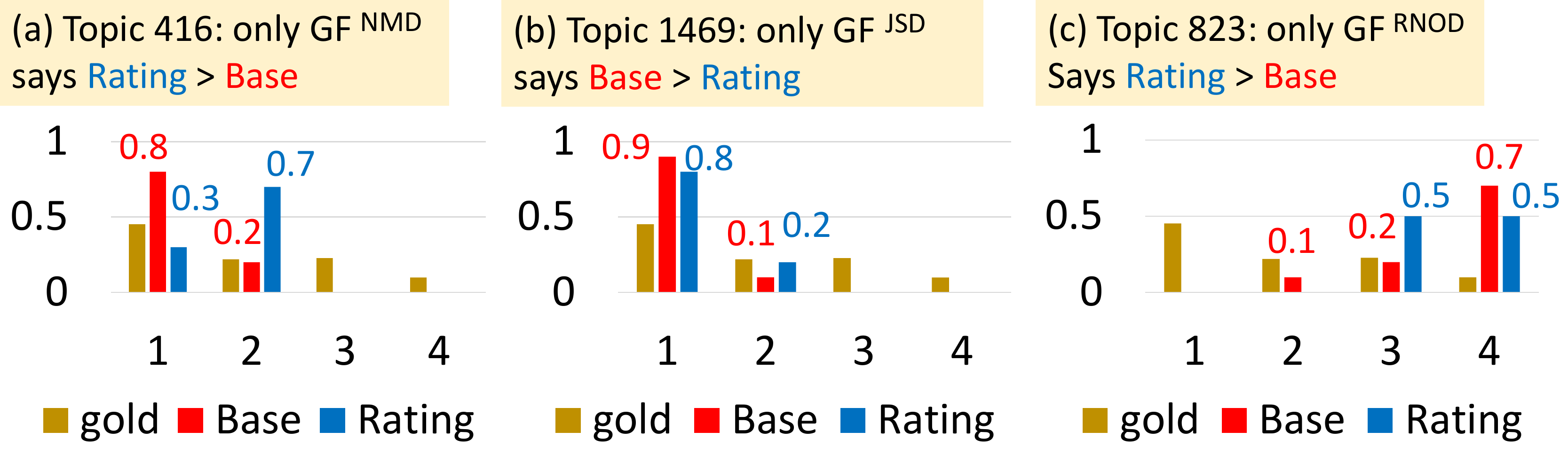}\vspace*{-2mm}
\caption{Review count distributions at rank $k=10$ achieved by Base and Rating for three example topics, with the common gold distribution.}\label{f:naver-dist}

\vspace*{-3mm}

\end{center}
\end{figure}

\begin{table}[t]
\centering

\caption{\textit{DistrSim} and revcnt-GF scores for the cases shown in Figure~\ref{f:naver-dist}.
Higher (i.e., better) scores are shown in bold.
}\label{t:naver-similarity}
\begin{scriptsize}
\begin{tabular}{c|c|c|c}
\hline
 				&\textit{DistrSim} / GF (JSD)	&\textit{DistrSim} / GF (NMD)	&\textit{DistrSim} / GF (RNOD)\\
\hline
\multicolumn{4}{c}{(a) Topic 416 (GF$^{\mathrm{NMD}}$ is the outlier)}\\
\hline
\textbf{Base}			&\textbf{0.801} / \textbf{0.572}		&0.742 / 0.566					&\textbf{0.712} / \textbf{0.505}\\
\textbf{Rating}			&0.730 / 0.366					&\textbf{0.807} / \textbf{0.576}		&0.668 / 0.365\\
\hline
\multicolumn{4}{c}{(b) Topic 1469 (GF$^{\mathrm{JSD}}$ is the outlier)}\\
\hline
\textbf{Base}			&0.765 / \textbf{0.554}				&0.708 / 0.553					&0.643 / 0.480\\
\textbf{Rating}			&\textbf{0.801} / 0.532				&\textbf{0.742} / \textbf{0.614}		&\textbf{0.712} / \textbf{0.512}\\			
\hline
\multicolumn{4}{c}{(c) Topic 823 (GF$^{\mathrm{RNOD}}$ is the outlier)}\\
\hline
\textbf{Base}			&\textbf{0.574} / \textbf{0.417}		&0.458 / \textbf{0.454}				&0.464 / 0.380\\
\textbf{Rating}			&0.521 / 0.408					&\textbf{0.492} / 0.423)				&\textbf{0.528} / \textbf{0.414}\\
\hline
\end{tabular}\vspace*{-2mm}
\end{scriptsize}

\end{table}

To investigate the differences in the \textit{DistrSim} functions and the revcnt-GF measures that are based on them,
we first filtered the 418 queries
and obtained those where one of the three GF measures disagreed with the other two;
there were 102 such queries.
Within this query set,
revcnt-GF$^{\mathrm{NMD}}$ disagreed with the other two for 83 queries (Case~A);
revcnt-GF$^{\mathrm{JSD}}$ disagreed with the other two for 16 queries (Case~B); and
revcnt-GF$^{\mathrm{RNOD}}$ disagreed with the other two for the remaining 3 queries (Case~C).
These results also show that revcnt-GF$^{\mathrm{NMD}}$
tends to be the outlier.
It is known that RNOD lies between
JSD and NMD in terms of how it behaves as a divergence measure~\cite{sakai21acl,sakai21cikmlq}:
our results show that the GF measures inherit their properties,
as the GF measures are essentially weighted averages of \textit{DistrSim} scores obtained at each rank (See Eq.~\ref{eq:GF}).

To examine how these discrepancies between the three revcnt-GF measures arise,
we selected one topic each from Cases~A, B, and C with relatively large score discrepancies, and examined the results closely.
%(We omit ``revnt-'' hereafter.)
Figure~\ref{f:naver-dist} visualises the distributions over the review count groups achieved at rank 10
by \textbf{Base} and \textbf{Rating} for the three selected topics, together with the common gold distribution.
Table~\ref{t:naver-similarity} 
shows the corresponding \textit{DistrSim} scores as well as the final revcnt-GF scores.
%Recall that the GF scores are basically weighted averages of \textit{DistrSim} scores (Eq~\ref{eq:GF}),
%where the weights are RBP-based in this particular experiment.
Table~\ref{t:naver-similarity}(a) shows that for Topic~416,
GF$^{\mathrm{NMD}}$ disagrees with GF$^{\mathrm{JSD}}$ and GF$^{\mathrm{RNOD}}$
precisely because $\textit{DistrSim}^{\mathrm{NMD}}$ disagrees with $\textit{DistrSim}^{\mathrm{JSD}}$
and $\textit{DistrSim}^{\mathrm{RNOD}}$.
However, Figure~\ref{f:naver-dist}(a) shows that  this behaviour of $\textit{DistrSim}^{\mathrm{NMD}}$
is rather counterintuitive:
since the gold distribution gives the highest probability to 
the zero-review group (Group~1),
the achieved distribution of \textbf{Base} (red) seems better than
that of \textbf{Rating} (blue).
On the other hand,
Table~\ref{t:naver-similarity}(b) shows that for Topic~1469,
while GF$^{\mathrm{JSD}}$ disagrees with the other two,
all three \textit{DistrSim} functions agree that
\textbf{Rating} is better than \textbf{Base} at rank 10.
(Figure~\ref{f:naver-dist}(b) shows the distributions.)
That is, this discrepancy at the GF score level is 
due to the
RBP-based
weighted averaging step  of Eq.~\ref{eq:GF}.
%, where the weights are RBP-based
%and top-heavy.
Finally, Table~\ref{t:naver-similarity}(c) 
shows that for Topic~823,
while GF$^{\mathrm{RNOD}}$ is the outlier at the GF score level,
both $\textit{DistrSim}^{\mathrm{NMD}}$ and $\textit{DistrSim}^{\mathrm{RNOD}}$
(i.e., those that can handle ordinal classes)
prefer \textbf{Rating} over \textbf{Base}.
That is, $\textit{DistrSim}^{\mathrm{JSD}}$ is the actual outlier at the \textit{DistrSim} level.
From Figure~\ref{f:naver-dist}(c),
it can be observed that, for this topic, the order-aware measures
penalise \textbf{Base} heavily for emphasing Group~4 (items with over 100 reviews) too much.
Based on the above analysis, our recommendation for handling ordinal attribute sets
is to use multiple similarity functions (e.g., $\textit{DistrSim}^{\mathrm{JSD}}$ and $\textit{DistrSim}^{\mathrm{RNOD}}$),
pay attention to cases where they disagree,
and, if possible, examine which function seems more intuitive.
As we have discussed in Section~\ref{s:intro},
researchers should also be aware that JSD ignores the ordinal nature of classes and therefore may not be appropriate for some applications.

\begin{figure}[t]
\begin{center}

\includegraphics[width=0.25\textwidth]{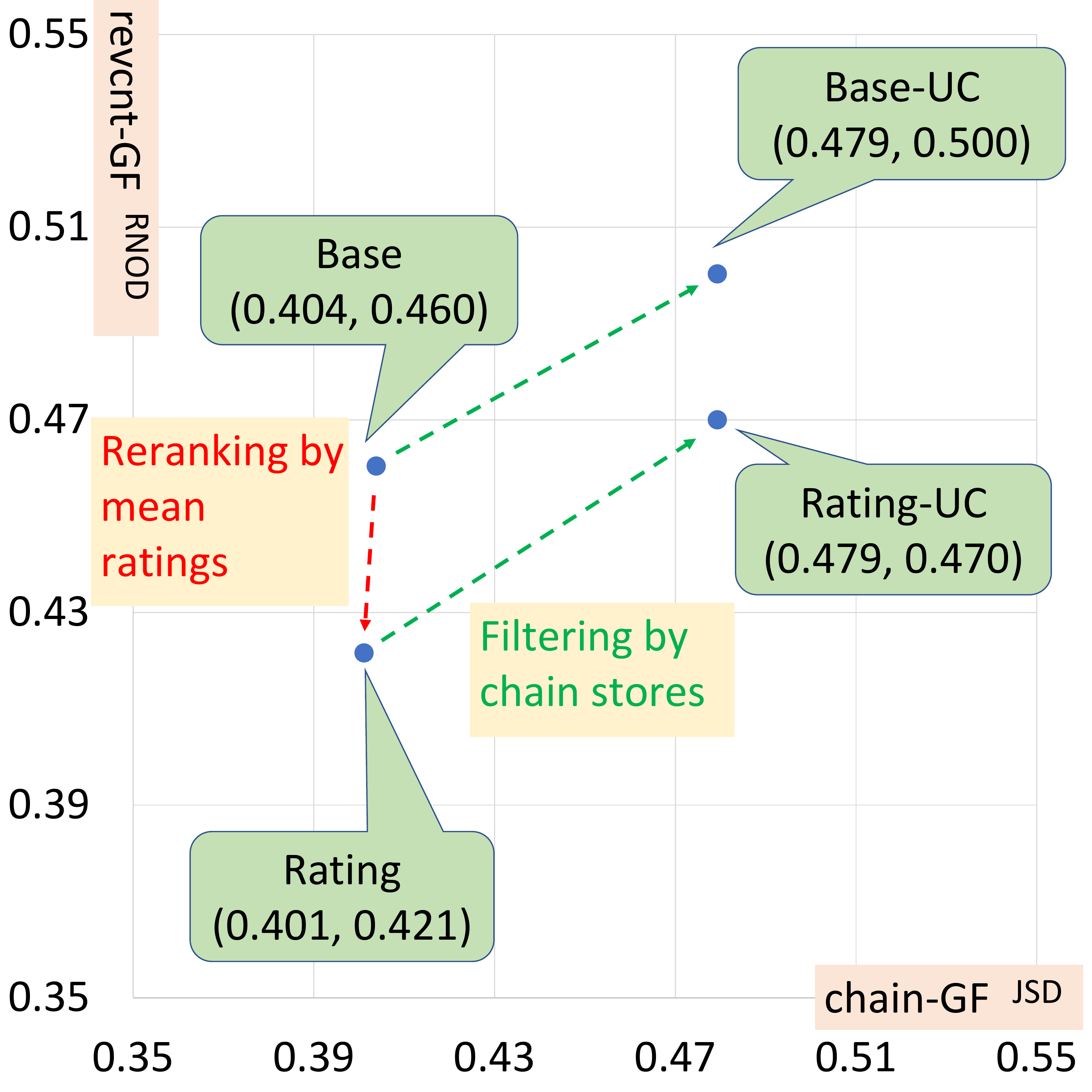}\vspace*{-2mm}
\caption{Visualising intersectional group fairness using GF measures for the local search runs.
}\label{f:naver-intersectional}

\vspace*{-2mm}

\end{center}
\end{figure}

Figure~\ref{f:naver-intersectional} provides a  visual summary of our Local Search experiment,
by plotting the revcnt-GF$^{\mathrm{RNOD}}$ scores against the
chain-GF$^{\mathrm{JSD}}$ scores for the 4 runs we considered.
Again, it is clear that the Unique Chain filtering step improves the ranking in terms of both
chain-based and revcnt-based group fairness (green arrows),
and that reranking by mean rating hurts the revcnt-based group fairness only (red arrow).
We have thus demonstrated that our framework enables researchers to study
intersectional group fairness, even when both nominal and ordinal attribute sets are involved.

We have also conducted a smaller experiment with both nominal and ordinal attribute sets
using data available from
\emph{Inside Airbnb}~\cite{biega18}.\footnote{
\url{http://insideairbnb.com/get-the-data.html} (visited September 8, 2021)
}
The results are similar to what we have reported here, and are omitted in this paper.
%due to lack of space.
%However, the results are omitted here due to lack of space.

\section{Conclusions}\label{s:conclusions}

We presented a simple and versatile framework for evaluating ranked lists in terms of group fairness and relevance,
where the groups can be either nominal or ordinal in nature.
First, we demonstrated that, if the attribute set is binary,
our framework can easily quantify the overall polarity of each ranked list.
Second, by utilising an existing diversified search test collection and treating each intent as an attribute value,
we demonstrated that our framework can handle soft group membership, and that
our group fairness measures are highly correlated with both adhoc IR and diversified IR measures under this setting.
Third, we demonstrated how our framework can quantify intersectional group fairness
based on multiple attribute sets. We also showed that the choice of the similarity function
for comparing the achieved and target distributions over the attribute values matters in some cases.
Our recommendation is to use multiple similarity functions (e.g., $\textit{DistrSim}^{\mathrm{JSD}}$ and $\textit{DistrSim}^{\mathrm{RNOD}}$)
if ordinal attribute values need to be considered.
Data that are necessary for reproducing our experimental results 
are available from \url{https://waseda.box.com/GFR20220401targz}\ .

Our future work includes further investigation of the properties of the similarity functions in the context of group-fair ranking evaluation,
and implementing this framework in a shared task.

%\clearpage

%To be disclosed upon acceptance

%\newpage

\clearpage

\bibliographystyle{ACM-Reference-Format}

%\begin{small}

\balance

\vspace*{3mm}\bibliography{sigir22reject}

%\end{small}

\end{document}